\documentclass[usenatbib,useAMS,1column]{mn2e}
\usepackage{natbib}
\usepackage[english]{babel}
\usepackage{graphicx,epsfig}
\usepackage{color,subfigure}
\usepackage{morefloats}
\usepackage{amsmath,amssymb}
\usepackage{lscape}
\usepackage{hyperref}
\usepackage{url}
\usepackage{fancyhdr}
\usepackage{supertabular}
\usepackage{multicol}
\usepackage{multirow}
\usepackage{longtable}
\usepackage[most]{tcolorbox}
\usepackage[justification=centering]{caption}
\usepackage{caption}
\usepackage{rotating}

\begin{document}

\title{Determining the torus covering factors for a sample of type~1 AGN in the local Universe}
\author[Ezhikode et. al. ]{Savithri H. Ezhikode,$^{1}$\thanks{Contact e-mail: savithrihezhikode@gmail.com} Poshak Gandhi,$^{2}$ Chris Done,$^{3}$ Martin Ward,$^{3}$
\newauthor
Gulab C. Dewangan,$^{4}$ Ranjeev Misra,$^{4}$ Ninan Sajeeth Philip$^{1}$\\
$^{1}$ Department Of Physics, St. Thomas College, Kozhencherry, Kerala 689641, India\\
$^{2}$ School of Physics \& Astronomy, University of Southampton, Highfield, Southampton SO17 1BJ, UK\\
$^{3}$ Centre for Extragalactic Astronomy, Department of Physics, Durham University, South Road, Durham DH1 3LE, UK\\
$^{4}$ Inter-University Centre for Astronomy \& Astrophysics, Post Bag 4, Ganeshkhind, Pune, India\\}

\maketitle

\begin{abstract}
{ 
In the unified scheme of active galactic nuclei, a dusty torus
absorbs and then reprocesses a fraction of the intrinsic luminosity which is emitted at longer wavelengths. Thus, subject to radiative transfer corrections, the fraction of the sky covered by the torus as seen from the central source (known as the covering factor $f_c$) can be estimated from the ratio of the infrared to the bolometric luminosities of the source as $f_c$=$L_{\rm torus}$/$L_{\rm Bol}$. However, the uncertainty in determining $L_{\rm Bol}$ has made the estimation of covering factors by this technique difficult, especially for AGN in the local Universe where the peak of the observed SEDs lies in the UV (ultraviolet). Here, we determine the covering factors of an X-ray/optically selected sample of 51 type~1 AGN. The bolometric luminosities of these sources are derived using a self-consistent, energy-conserving model that estimates the contribution in the unobservable far-UV region, using multi-frequency data obtained from SDSS, \textit{XMM-Newton}, \textit{WISE}, 2MASS and UKIDSS. We derive a mean value of $f_c$\,$\sim$\,0.30 with a dispersion of 0.17. Sample correlations, combined with simulations, show that $f_c$ is more strongly anti-correlated with $\lambda_{\rm Edd}$ than with $L_{\rm Bol}$. This points to large-scale torus geometry changes associated with the Eddington-dependent accretion flow, rather than a receding torus, with its inner sublimation radius determined solely by heating from the central source. Furthermore, we do not see any significant change in the distribution of $f_c$ for sub-samples of radio-loud sources or Narrow Line Seyfert~1 galaxies (NLS1s), though these sub-samples are small.
}
\end{abstract}

\begin{keywords}
galaxies:active $-$ galaxies:Seyfert $-$ X-rays:galaxies $-$ ultraviolet: galaxies $-$ infrared: galaxies		
\end{keywords}

\section{Introduction}
\label{Sec1}

Studying broadband spectral energy distributions (SEDs) of active galactic nuclei (AGN) can shed light on the emission mechanisms operating in the distinct physical components of the AGN. For example, the big blue bump (BBB) \citep{1980ApJ...235..361R, 1989ApJ...347...29S, 1994ApJS...95....1E} in the optical/UV band is associated with modified black body emission from the accretion disc, whereas the Comptonised emission from the corona \citep{1993ApJ...413..507H} produces an X-ray power law continuum above $\sim$~2~keV. Even though the spectral features of various classes of AGN are distinct, the so-called unification scheme \citep[for reviews see][]{1993ARA&A..31..473A, 1995PASP..107..803U, 2015ARA&A..53..365N} postulates that the different types of AGN are intrinsically similar. The model suggests the presence of a dusty, molecular torus shaped structure, surrounding the central source, which gives rise to anisotropic emission in polar directions. The observed characteristics of AGN are governed by the orientation of this obscuring torus with respect to our line of sight. 

This torus is optically-thick, with a size of 0.1$-$10~pc \citep{2006ApJ...639...46S, 2007A&A...476..713K, 2013A&A...558A.149B}. The region has a gas density in the range of 10$^4-$10$^7$~cm$^{-3}$ while the column density ranges from $\sim10^{22}$ to $\sim10^{25}$~cm$^{-2}$ \citep{2013peag.book.....N}. In the most simple orientation-based unification scheme, the broad-line region may or may not be obscured by the torus material depending on the inclination angle of the system with respect to our line of sight. In this picture, the classification as a type~2 or a type~1 AGN is determined by the orientation alone. The obscuration is parameterised by the opening angle of the torus which in turn determines the covering factor $f_c$. Hence the covering factor is defined as the fraction of the sky that the torus blocks/absorbs the emission from the central source. 

The torus is directly exposed to the emission from the central engine and the photons illuminating this region are absorbed by the dust grains. The heated dust then re-radiates these absorbed optical/UV photons in the infrared (IR) band. Hence the IR continuum is attributed to the thermal emission from the silicate and graphite grains with a broad temperature distribution extending up to the sublimation temperatures of about 1500~K. Cooler dust on larger scales emits at longer wavelengths and also shows a silicate feature around $\sim$~10~$\mu$m. 

Dust grains can no longer survive if heated above their sublimation temperature. As a result, the inner radius of the dust is determined by the distance from the centre, at which the dust is sublimated by the primary continuum. A more luminous AGN heats the dust more strongly and hence the sublimation radius will be larger, with $R\propto L_{\rm Bol}^{1/2}$ \citep{1987ApJ...320..537B}. If the dust distribution has a fixed scale height (as opposed to scaling with mass and/or mass accretion rate), then this means that the covering factor of the dusty torus decreases as the luminosity of the source increases. This anti-correlation between $f_c$ and the luminosity was suggested by \cite{1991MNRAS.252..586L} and is called the \textit{receding torus model}. This does not depend on the detailed dust distribution. For example a clumpy torus (\citealt{2008ApJ...685..160N, 2010A&A...523A..27H}) would exhibit much the same behaviour as the covering factor is determined mostly by the total solid angle covered by the dust. Similarly, if most of the mid-infrared (MIR) emission arises from scales beyond the classical torus \citep{2013ApJ...771...87H,2016A&A...591A..47L,2016ApJ...822..109A}, then $f_c$ would be a measure of the efficiency of this extended emission component and how it scales with AGN power.

There are various methods for determining $f_c$. One is a statistical approach, 
based on optical demographic studies of AGN which compared the fraction of type 1 and 2 AGN, with $L_{[OIII]}$ used as a proxy for $L_{\rm Bol}$. Alternatively, this can be done in X$-$rays, using the fraction of X$-$ray unobscured to obscured AGN, with $L_{\rm X-ray}$ tracing $L_{\rm Bol}$. Both types of studies generally find a decrease in the fraction of obscured AGN with increasing luminosity, consistent with the receding torus model \citep[e.g. ][]{2005MNRAS.360..565S,2005AJ....129.1795H,2003ApJ...598..886U,2003ApJ...596L..23S,2005ApJ...635..864L,
2006ApJ...652L..79T,2008A&A...490..905H,2013PASJ...65..113T,2016ApJ...819..166M} but with a few exceptions \citep{2006MNRAS.372.1755D,2006ApJS..165...19E} or additional correlations \citep{2008A&A...490..905H}. However, neither the X$-$rays nor the optical emission lines, used as proxies, give a reliable estimate of the bolometric luminosity, $L_{\rm Bol}$ \citep[][hereafter J12a, J12b]{2007MNRAS.381.1235V,2013ApJ...777...86L,2012MNRAS.420.1825J,2012MNRAS.425..907J}, which is the key driving parameter in the \textit{receding torus model}, and there are also multiple selection effects \citep{2010ApJ...714..561L}.

Alternatively, the SED can be fitted over as wide a bandpass as possible to directly constrain $L_{\rm Bol}$ from  observations, and then the ratio of the IR luminosity to the bolometric luminosity $L_{\rm IR}/L_{\rm Bol}$ can be used to derive $f_c$ based on dust (re)emission. Again these studies generally show an anti-correlation with $L_{\rm Bol}$ \citep{2007ApJ...661...30G,2007A&A...468..979M,2008MNRAS.386.1252H,
2009MNRAS.399.1206H,2009A&A...502..457G,2013MNRAS.429.1494R,2013ApJ...777...86L}. 
Nonetheless, there are still uncertainties. A major difficulty in the SED-based analysis is determining the bolometric luminosity of the AGN. A substantial part of the AGN luminosity emerges in the UV region and is unobservable due to the interstellar absorption with our Galaxy. Furthermore, the IR luminosity can be self-obscured, with radiation transfer effects through optically-thick dust affecting the IR luminosity observed for type~2 (obscured) objects \citep{2016MNRAS.458.2288S, 2008ApJ...679..140T,1992ApJ...401...99P}. 

In this work, we use a sample of 51 unobscured AGN, so these should not be affected 
by radiative transfer effects. They all have well sampled broadband optical-UV-X$-$ray data to define the SED. Additionally, they have $L_{\rm Bol}$ estimated by fits using a self-consistent energy-conserving model to bridge across the unobservable far-UV (FUV) region (J12a, J12b). In this paper, we extend these SEDs to include far-infrared (FIR) wavelengths to estimate the covering factors of the AGN in our sample. We then investigate the dependence of $f_c$ with different AGN properties.

This paper is organised as follows. In Section~\ref{Sec2}, we discuss the sample selection and preparation of the multi-wavelength data. Section~\ref{Sec3} gives a detailed description of the modelling of the broadband SED of the sample. Section~\ref{Sec4} deals with the main results obtained in the work. Section~\ref{Sec5} is dedicated to the discussion of the results and the conclusions are given in Section~\ref{Sec6}. The details regarding the local models used in this work, the broadband SED plot for each source, notes on selected sources, and other relevant information is given in the Appendix. Throughout this paper, we have adopted a cosmology with Hubble constant of H$_0$~=~70~km~s$^{-1}$~Mpc$^{-1}$, $\Omega_{\Lambda}$=0.73 and $\Omega_M$=0.27.

\section{Sample Selection \& Data Preparation}
\label{Sec2}

For our study, we choose the sample of 51 type~1 AGN analysed by J12a and J12b. These are selected to have good SDSS spectra (DR7) with $z<0.4$ so that H$\alpha$ and H$\beta$ lines (black hole mass estimator) are included in the bandpass, and good quality \textit{XMM-Newton} X-ray data are available, without complex absorption features. Optical band continuum points were extracted from the SDSS data by removing line emission, Balmer continuum, and host galaxy contribution. The \textit{XMM-Newton} satellite also has the Optical Monitor (OM) which provides simultaneous optical-UV photometry. These photometric points were extracted using 6\arcsec diameter apertures to minimise host galaxy contamination. Therefore these sources all have well sampled SEDs, which can be modelled to give a good estimate of their bolometric luminosity. The sample spans a broad range of AGN types comprising 12 Narrow Line Seyfert~1 galaxies (NLS1s), 39 Broad Line Seyfert~1 galaxies (BLS1s), a broad absorption line (BAL) quasar which is also radio-loud (PG~1004+130, No.~13) and two more radio-loud AGN RBS~0875 (No.~14) and PG~1512+370 (No.~45). Further information of these sources is given in Table~1 of J12a.

\begin{table}[]
\scriptsize
\tabcolsep 3.5pt
\centering
\caption{The bandwidth in $\mu$m (Col.~3), the effective wavelength $\lambda_{\rm eff}$ in $\mu$m (Col.~4), the zero-point flux density $F_{\nu0}$ in Jy (Col.~5) to convert from magnitudes and the extinction correction factor (Col.~6) for \textit{WISE}, 2MASS \& UKIDSS bands.}
\begin{tabular}{cccccc}
\hline
\hline
\\
\multicolumn{1}{c}{\multirow{2}{*}{Survey}} & \multirow{2}{*}{Band} & Bandwidth & \multicolumn{1}{c}{$\lambda_{\rm eff}$} & \multicolumn{1}{c}{$F_{\nu0}$} & \multicolumn{1}{c}{\multirow{2}{*}{$A_{\lambda}/A_V$}} \\\\
\multicolumn{1}{c}{}  &  & $\mu$m & \multicolumn{1}{c}{$\mu$m} & \multicolumn{1}{c}{Jy} & \multicolumn{1}{c}{}                        \\
\hline
\multicolumn{1}{c}{\multirow{4}{*}{\textit{WISE}}}  &  \textit{W}1  &  0.663$\pm$0.001  &  3.35$\pm$0.01  &  309.5$\pm$4.6  &  0.069  \\

\multicolumn{1}{c}{}  &  \textit{W}2  &  1.042$\pm$0.001  &  4.60$\pm$0.02 &  171.8$\pm$2.5  &  0.053  \\

\multicolumn{1}{c}{}  &  \textit{W}3  &  5.510$\pm$0.020  &  11.56$\pm$0.04 &  31.7$\pm$0.5   &  0.068  \\

\multicolumn{1}{c}{}  &  \textit{W}4  &  4.100$\pm$0.040  &  22.09$\pm$0.12 &  8.4$\pm$0.3  &  0.052  \\
\hline
\multirow{3}{*}{2MASS}  &  J  &  0.162$\pm$0.001  &  1.235$\pm$0.006  &  1594.0$\pm$27.8  &  0.282  \\
                        &  H  &  0.251$\pm$0.002  &  1.662$\pm$0.009  &  1024.0$\pm$20.0  &  0.190  \\                                                                   
                        &  K  &  0.262$\pm$0.002  &  2.159$\pm$0.011  &  666.7$\pm$12.6 &  0.114  \\
\hline
\multirow{4}{*}{UKIDSS} &  Y  &  0.102  &  1.031  &  2026.0  &  0.380  \\
                        &  J  &  0.159  &  1.248  &  1530.0  &  0.282  \\
                        &  H  &  0.292  &  1.631  &  1019.0  &  0.190  \\
                        &  K  &  0.351  &  2.201  &  631.0  &  0.114   \\
\hline    
\hline                                   
\label{tab_IRband}
\end{tabular}
\end{table}

\begin{figure}
\begin{center}
\includegraphics[trim=0cm 0cm 0cm 0.0cm, clip=true, width =8.5cm, angle=0]{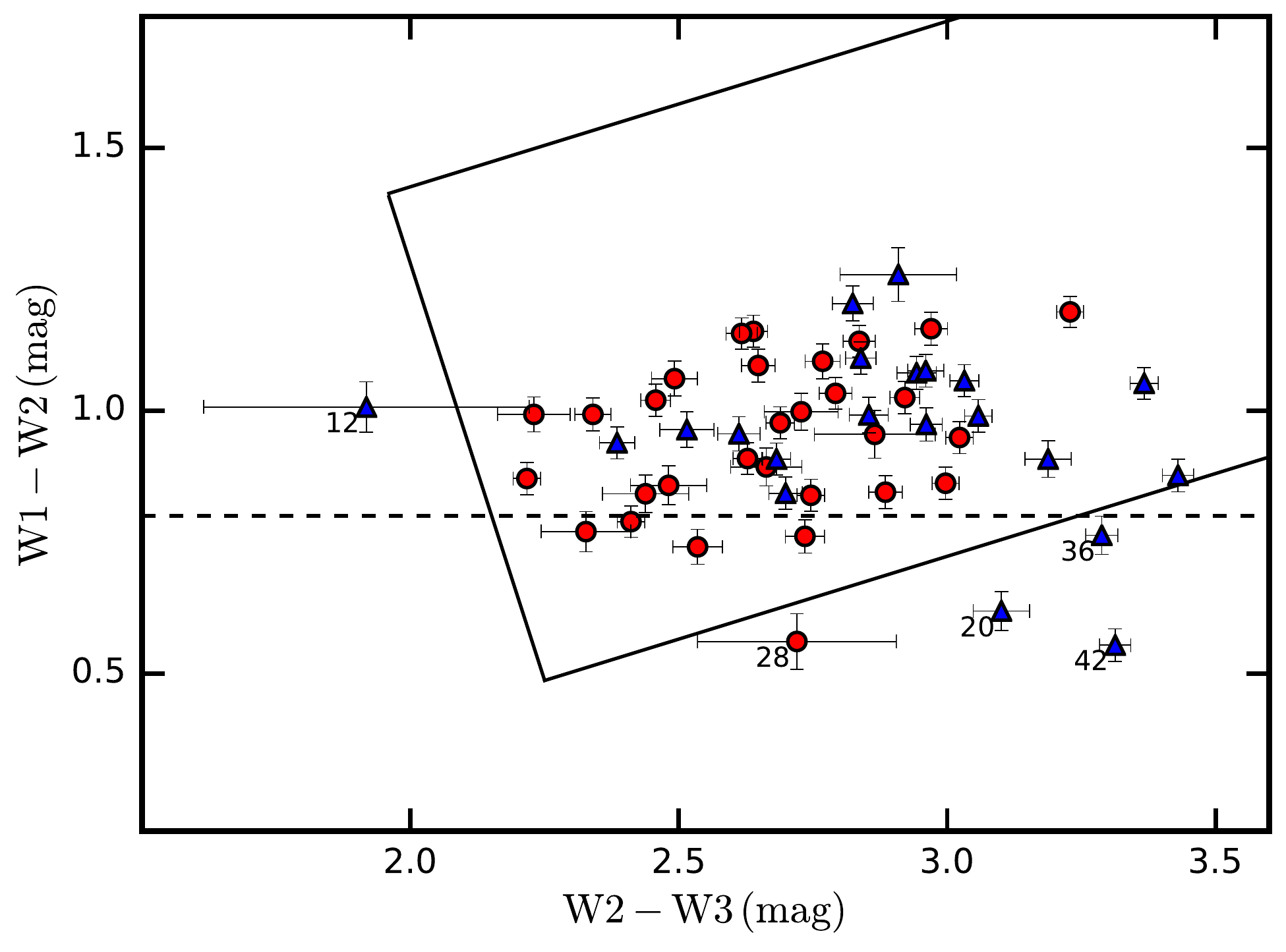}
\caption{\small \textit{WISE} colour-colour plot for our sample. The solid line describes the AGN wedge \citep{2012MNRAS.426.3271M} and the colour-cut \citep{2012ApJ...753...30S} is plotted with the dashed line. Blue triangles denote objects with significant host galaxy contribution ($L_{\rm host} > 10\%$ of $L_{\rm Bol}$) in the model fit.}
\label{Fig_AGNwedge}
\end{center}
\end{figure}

\subsection{IR Data}
\label{IRdata}

We extended the continuum of the SEDs discussed in J12b to include IR data obtained from the \textit{Wide-field Infrared Survey Explorer} (\textit{WISE}; \citealt{2010AJ....140.1868W}), Two Micron All-Sky Survey (2MASS; \citealt{2006AJ....131.1163S}) and UKIRT Infrared Deep Sky Survey (UKIDSS) catalogues \citep{2006MNRAS.367..454H}. These span a wavelength range from $\sim$\,1~$\mu$m to 20~$\mu$m, giving good coverage from near-infrared (NIR) to mid-infrared wavelengths. For cases where both 2MASS and UKIDSS data were available for each source, we opted to use the UKIDSS data for the NIR band due to its smaller aperture size. Our principal results, and specifically the distribution of covering factors, do not depend on the choice of 2MASS or UKIDSS data, a point which is expanded on in Appendix~\ref{E}. We used 2MASS only for sources for which there are no UKIDSS observations. One object in the sample (2XMM~J100523.9+410746; No.~12) does not have either UKIDSS or 2MASS data. In that case, we could use only the \textit{WISE} data for the IR analysis. The bandpasses with their zero magnitude flux and aperture sizes are given in Table~\ref{tab_IRband}, together with the extinction corrections in each band, $A_{\lambda}$, extracted from \citet{1989ApJ...345..245C} with $R_V$=3.1 (see \cite{2011ApJ...740L..13G} and references therein for \textit{WISE} extinction corrections). The resulting NIR-MIR fluxes, corrected for Galactic reddening, are listed in Table~\ref{tab_IRflux}. We incorporate these into {\sc xspec} using the {\sc ftool} {\sc flx2xsp}. 

We investigated the dominant IR emission mechanism based on \textit{WISE} colour selection thresholds of \cite{2012MNRAS.426.3271M} and \cite{2012ApJ...753...30S}. 
The MIR colour-cut defined by \cite{2012ApJ...753...30S} identifies the AGN candidates with \textit{W}1-\textit{W}2~$\geq$~0.8. In addition, the AGN wedge of \citet{2012MNRAS.426.3271M} is designed to select objects with red MIR power-law SEDs in the first three bands of \textit{WISE}. Fig.~\ref{Fig_AGNwedge} shows that most of the sources in our sample, except 2XMM~J100523.9+410746 (No.~12), RX~J1140.1+0307 (No.~20), RX~J1233.9+0747 (No.~28), 1E~1346+26.7 (No.~36) \& NGC~5683 (No.~42), are within the AGN wedge and above the colour-cut. This confirms that the MIR is likely dominated by the AGN rather than the host galaxy in most cases. The MIR fluxes for those which lie below the colour wedge and the colour-cut are likely to be dominated by the stellar population or star formation activity in the host galaxy, especially in the case of \textit{W}1 filter. For example, the MIR SED of NGC~5683 (No.~42) shows a significant contribution from the host galaxy.

\section{The Broadband SED Model}
\label{Sec3}

Multiwavelength observations are a crucial ingredient in understanding the physical processes occurring in AGN and to study the structure of their inner regions. Some notable features in the broadband SED model of AGN are; the hard X-ray power law, the soft X-ray excess below 2 keV, the big blue bump which peaks in optical/UV region, and the infrared bump at $\sim$10~$\mu$m. The optical/UV emission in AGN is thought to arise from a multi-temperature accretion disc. The power-law component originates from the inverse Compton scattering of accretion disc photons by a hot, optically-thin corona. The infrared emission results from reprocessing of the absorbed optical/UV/X-ray emission from the AGN.

\subsection{Modelling the Optical/UV \& X$-$rays}

The SED can be phenomenologically fitted by a black body component for the accretion disc and thermal Comptonisation from an optically-thin, high-temperature corona to model the hard X-ray power law above 2 keV. At lower wavelengths, the soft X-ray excess can be modelled with an optically-thick, low-temperature thermal Comptonisation model. In J12b, they modelled the Optical/UV \& X$-$ray continua with the {\sc XSPEC} model {\sc optxagnf}
\footnote{A description of {\sc optxagnf} can be found in the {\sc XSPEC} website \url{http://heasarc.nasa.gov/xanadu/xspec/models/optxagn.html}} \citep{2012MNRAS.420.1848D}. This model is fully self-consistent (energy conserving), and associates all components with emission from the accretion disc itself and energy extracted from it i.e. the soft X-ray component and the hard X-ray power-law. {\sc optxagnf} is parameterised by black hole mass (M$_{\rm BH}$), Eddington ratio (\textbf{$L/L_{\rm Edd}$ or $\lambda_{\rm Edd}$}), black hole spin ($\sc a$), coronal radius ($R_{\rm cor}$), outer radius of the accretion disc ($R_{\rm out}$), electron temperature ($kT_{\rm e}$) and optical depth ($\tau$) of the corona producing soft X-ray component, hard X-ray (2$-$10~keV) photon index ($\Gamma$), fraction of the coronal energy emitted in the hard X-ray power law ($f_{\rm pl}$) and redshift ($z$) i.e., ten parameters in total. In detail, this model introduces $R_{\rm cor}$, the radius down to which the gravitational energy is released as black body emission in the disc. Within this radius, the energy is emitted as the soft X-ray excess and the high energy power law. In this model, the mass accretion rate is constrained by the optical/UV luminosity. If the black hole mass is known, then these parameters can be used to estimate the total luminosity. J12b assumed an accretion efficiency of 0.057 for a Schwarzschild black hole, and hence determined the total luminosity. The study presented in J12b is a refinement to the SED fitting given in J12a. This now includes a self-consistent colour temperature correction for the standard disc emission.

We adopted the same procedures as described in J12b, and performed the spectral analysis of the optical/UV and X-ray continua using {\sc optxagnf} in {\sc XSPEC} version 12.8.2. We model the Galactic absorption and reddening using the standard routines \textit{wabs} and \textit{redden} respectively, and any intrinsic absorption/reddening using \textit{zwabs}/\textit{zredden}. The intrinsic column density was left free to vary during the spectral fitting while the Galactic column density for each source was frozen to the value obtained from LAB Survey \citep{2005A&A...440..775K}. We fixed the black hole mass to the best-fit value obtained by J12b in their models, who fixed upper(lower) limits to the mass from the broad(intermediate) velocity width of the H$\beta$ line decomposition. We also follow the method described in J12a, J12b and fix the outer radius of the accretion disc to be $\rm 10^4 R_g$. This is probably an upper limit to the size of the disc, as both the self-gravity radius and the best-fit to the disc emission generally indicate a somewhat smaller disc \citep{2010ApJ...724L..59H,2016arXiv161004221C}. The redshift\footnote{Redshift for each source is taken from \url{https://ned.ipac.caltech.edu/}} $z$ is fixed to the value given in Table~\ref{tab_IRflux}, this leaves 9 free parameters for the X-ray fitting portion of the model.

\subsection{Modelling the Infrared}

Our extended wavelength IR coverage is modelled as a combination of dust re-radiation and host galaxy emission. For the dust, we use the Seyfert~1 (unobscured) template of \cite{2004MNRAS.355..973S}, and for the host galaxy we use a range of 13 templates spanning ellipticals, spirals, and star-forming galaxies, from the SWIRE library \citep{2007ApJ...663...81P}. We incorporate these into {\sc xspec} as local models, which we call \textit{agndust} and \textit{hostpol}. Further details of \textit{agndust} and \textit{hostpol} templates are given in APPENDIX~\ref{A}. We use only the Seyfert~1 dust template in our fits since all our objects are unobscured, but we do investigate the entire range of host galaxy templates, and then adopt the one which gives the best-fit to each object.\\
\\
The optical/UV, X-ray and IR continua of each source in the sample were fitted by the final model $constant$($hostpol+agndust+wabs \times redden \times zwabs \times zredden \times$ {\sc optxagnf}), where the \textit{constant} only allows for small cross-calibration differences in normalisation between the X-ray spectra obtained from the three independent cameras EPIC-pn, MOS1 and MOS2, aboard the \textit{XMM-Newton} satellite. The broadband SED model for each source has 11 free parameters (including the normalizations of \textit{agndust} and \textit{hostpol}), apart from the multiplicative factor of the model \textit{constant}. The fit-statistic and the best-fit hostpol template for each source are given in Table~\ref{tab_parameter}. The plots showing the data and individual model components for all the sources are given in the APPENDIX~\ref{B}. Mrk~0110 (No.~9) shows a clear discrepancy in the SDSS data due to extreme variability, and therefore has a very large $\chi^2$. Also, there are two super-Eddington sources in our sample, KUG~1031+398 (No.~15) and PG~2233+134 (No.~50). ​These sources are discussed individually in APPENDIX~\ref{C}. We note that the potential discrepancies mentioned above do not influence any of our resulting inferences on the $f_c$ distribution.

\begin{landscape}
\begin{table}
\tabcolsep 3.0pt
\scriptsize
\begin{center}
\caption{Dereddened IR flux for the sample in each band of \textit{WISE}, 2MASS, and UKIDSS.}
\label{tab_IRflux}
\begin{tabular}{ccclllllllllll}
\hline
\hline
\multirow{3}{*}{No.}   & \multirow{3}{*}{Object}    & \multirow{3}{*}{Redshift}        & \multicolumn{11}{c}{IR flux (mJy)}                                       \\               \\ \cline{4-14} 

                     &                            &                           & \multicolumn{4}{c}{} & \multicolumn{3}{c}{} & \multicolumn{4}{c}{} \\ 

                      &                            &                           & \multicolumn{4}{c}{\textit{WISE}} & \multicolumn{3}{c}{2MASS} & \multicolumn{4}{c}{UKIDSS} \\ \cline{4-14} 

                      &                            &                           &      &      &     &      &        &        &        &       &      &      &      \\ 

                      &                            &                           & \textit{W}1   & \textit{W}2   & \textit{W}3   & \textit{W}4  & \textit{J}       & \textit{H}      & $K_s$      & \textit{Y}     & \textit{J}    & \textit{H}    & \textit{K}    \\ \hline

\multicolumn{1}{l}{1} & \multicolumn{1}{l}{UM~269}       & \multicolumn{1}{l}{0.308}      &  2.10$\pm$0.06      &  2.90$\pm$0.07      &  4.22$\pm$0.26     &  7.94$\pm$1.37     &  0.45$\pm$0.05         &  0.70$\pm$0.08        &  0.93$\pm$0.07        &  -       &  -      &  -      &  -     \\ 

\multicolumn{1}{l}{2} & \multicolumn{1}{l}{Mrk~1018}       & \multicolumn{1}{l}{0.043}      &  26.58$\pm$0.67      &  30.51$\pm$0.72      &  52.38$\pm$1.07     &  85.36$\pm$3.86     &  8.34$\pm$0.58         &  11.73$\pm$0.99        &  15.35$\pm$0.75        &  -       &  -      &  -      &  -     \\ 

\multicolumn{1}{l}{3} & \multicolumn{1}{l}{NVSS~J030639}       & \multicolumn{1}{l}{0.107}      &  5.15$\pm$0.13      &  6.22$\pm$0.15      &  16.52$\pm$0.42     &  41.50$\pm$2.28     &  -         &  -        &  -        &  1.02$\pm$0.03       &  1.34$\pm$0.04      &  1.99$\pm$0.06      &  2.97$\pm$0.09     \\ 

\multicolumn{1}{l}{4} & \multicolumn{1}{l}{2XMM~J074601.2+280732}       & \multicolumn{1}{l}{0.145}      &  1.35$\pm$0.04      &  1.53$\pm$0.04      &  2.43$\pm$0.18     &  3.84$\pm$1.19     &  0.52$\pm$0.06         &  0.66$\pm$0.09        &  0.79$\pm$0.08        &  -       &  -      &  -      &  -     \\ 

\multicolumn{1}{l}{5} & \multicolumn{1}{l}{2XMM~J080608.0+244421}       & \multicolumn{1}{l}{0.358}      &  1.12$\pm$0.03      &  1.56$\pm$0.04      &  3.58$\pm$0.22     &  8.00$\pm$1.28     &  -         &  -        &  -        &  0.25$\pm$0.01       &  0.30$\pm$0.01      &  0.40$\pm$0.01      &  0.56$\pm$0.02     \\ 

\multicolumn{1}{l}{6} & \multicolumn{1}{l}{HS~0810+5157}       & \multicolumn{1}{l}{0.377}      &  1.89$\pm$0.05      &  3.18$\pm$0.08      &  7.99$\pm$0.25     &  21.07$\pm$1.51     &  0.63$\pm$0.06         &  0.87$\pm$0.09        &  1.29$\pm$0.09        &  -       &  -      &  -      &  -     \\ 

\multicolumn{1}{l}{7} & \multicolumn{1}{l}{RBS~0769}       & \multicolumn{1}{l}{0.160}      &  3.33$\pm$0.09      &  5.36$\pm$0.13      &  15.40$\pm$0.39     &  32.69$\pm$1.66     &  0.90$\pm$0.06         &  1.27$\pm$0.11        &  1.82$\pm$0.09        &  -       &  -      &  -      &  -     \\ 

\multicolumn{1}{l}{8} & \multicolumn{1}{l}{RBS~0770}       & \multicolumn{1}{l}{0.033}      &  37.65$\pm$0.97      &  45.26$\pm$1.07      &  105.76$\pm$2.17     &  211.87$\pm$8.67     &  10.03$\pm$0.46         &  13.40$\pm$0.73        &  20.01$\pm$0.75        &  -       &  -      &  -      &  -     \\ 

\multicolumn{1}{l}{9} & \multicolumn{1}{l}{Mrk~0110}       & \multicolumn{1}{l}{0.035}      &  21.67$\pm$0.56      &  29.57$\pm$0.70      &  65.58$\pm$1.39     &  109.36$\pm$4.88     &  4.57$\pm$0.23         &  5.65$\pm$0.32        &  8.58$\pm$0.33        &  -       &  -      &  -      &  -     \\ 

\multicolumn{1}{l}{10} & \multicolumn{1}{l}{PG~0947+396}       & \multicolumn{1}{l}{0.206}      &  7.93$\pm$0.20      &  12.71$\pm$0.3      &  26.91$\pm$0.57     &  55.27$\pm$2.77     &  1.99$\pm$0.08         &  2.75$\pm$0.11        &  5.34$\pm$0.17        &  -       &  -      &  -      &  -     \\ 

\multicolumn{1}{l}{11} & \multicolumn{1}{l}{2XMM~J100025.2+015852}       & \multicolumn{1}{l}{0.373}      &  0.30$\pm$0.01      &  0.54$\pm$0.02      &  1.47$\pm$0.14     &  4.46$\pm$1.05     &  -         &  -        &  -        &  0.11$\pm$0.01       &  0.11$\pm$0.01      &  0.16$\pm$0.01      &  0.21$\pm$0.01     \\ 

\multicolumn{1}{l}{12} & \multicolumn{1}{l}{2XMM J100523.9+410746}       & \multicolumn{1}{l}{0.206}      &  0.35$\pm$0.01      &  0.50$\pm$0.02      &  0.54$\pm$0.15     &  3.55$\pm$0.12     &  -         &  -        &  -        &  -       &  -      &  -      &  -     \\

\multicolumn{1}{l}{13} & \multicolumn{1}{l}{PG 1004+130}       & \multicolumn{1}{l}{0.241}      &  8.10$\pm$0.21      &  11.90$\pm$0.28      &  36.20$\pm$0.79     &  83.73$\pm$4.04     &  3.41$\pm$0.12         &  3.54$\pm$0.16        &  5.00$\pm$0.18        &  -       &  -      &  -      &  -     \\ 

\multicolumn{1}{l}{14} & \multicolumn{1}{l}{RBS~0875}       & \multicolumn{1}{l}{0.178}      &  6.72$\pm$0.17      &  8.86$\pm$0.21      &  14.84$\pm$0.41     &  29.33$\pm$1.9     &  1.70$\pm$0.10         &  2.21$\pm$0.15        &  3.58$\pm$0.15        &  -       &  -      &  -      &  -     \\ 

\multicolumn{1}{l}{15} & \multicolumn{1}{l}{KUG~1031+398}       & \multicolumn{1}{l}{0.043}      &  11.66$\pm$0.29      &  19.32$\pm$0.45      &  70.46$\pm$1.40     &  132.33$\pm$5.69     &  3.57$\pm$0.14         &  4.38$\pm$0.21        &  5.86$\pm$0.22        &  -       &  -      &  -      &  -     \\ 

\multicolumn{1}{l}{16} & \multicolumn{1}{l}{PG~1048+342}       & \multicolumn{1}{l}{0.167}      &  3.89$\pm$0.10      &  5.58$\pm$0.14      &  13.61$\pm$0.34     &  27.78$\pm$1.53     &  1.28$\pm$0.06         &  1.57$\pm$0.08        &  2.46$\pm$0.11        &  -       &  -      &  -      &  -     \\ 

\multicolumn{1}{l}{17} & \multicolumn{1}{l}{1RXS~J111007}       & \multicolumn{1}{l}{0.262}      &  2.07$\pm$0.05      &  3.08$\pm$0.07      &  8.62$\pm$0.27     &  22.14$\pm$1.33     &  0.33$\pm$0.05         &  0.71$\pm$0.07      &  1.16$\pm$0.09        &  -       &  -      &  -      &  -     \\ 

\multicolumn{1}{l}{18} & \multicolumn{1}{l}{PG~1115+407}       & \multicolumn{1}{l}{0.155}      &  9.00$\pm$0.23      &  11.54$\pm$0.27      &  24.18$\pm$0.53     &  49.53$\pm$2.36     &  2.21$\pm$0.12         &  3.14$\pm$0.18        &  5.36$\pm$0.20        &  -       &  -      &  -      &  -     \\ 

\multicolumn{1}{l}{19} & \multicolumn{1}{l}{2XMM~J112328.0+052823}       & \multicolumn{1}{l}{0.101}      &  2.64$\pm$0.07      &  3.65$\pm$0.09      &  9.43$\pm$0.28     &  16.42$\pm$1.30     &  -         &  -        &  -        &  0.49$\pm$0.02       &  0.64$\pm$0.02      &  0.96$\pm$0.03      &  1.55$\pm$0.05     \\ 

\multicolumn{1}{l}{20} & \multicolumn{1}{l}{RX~J1140.1+0307}       & \multicolumn{1}{l}{0.081}      &  1.35$\pm$0.04      &  1.32$\pm$0.04      &  4.28$\pm$0.19     &  12.06$\pm$1.35     &  -         &  -        &  -        &  0.35$\pm$0.01       &  0.45$\pm$0.01      &  0.65$\pm$0.02      &  -     \\ 

\multicolumn{1}{l}{21} & \multicolumn{1}{l}{PG~1202+281}       & \multicolumn{1}{l}{0.165}      &  7.41$\pm$0.19      &  10.57$\pm$0.25      &  29.02$\pm$0.65     &  87.03$\pm$3.74     &  2.03$\pm$0.08         &  2.53$\pm$0.10        &  4.82$\pm$0.15        &  -       &  -      &  -      &  -     \\ 

\multicolumn{1}{l}{22} & \multicolumn{1}{l}{1AXG~J121359+1404}       & \multicolumn{1}{l}{0.154}      &  1.06$\pm$0.03      &  1.34$\pm$0.04      &  2.89$\pm$0.17     &  7.57$\pm$1.08     &  -         &  -        &  -        &  0.36$\pm$0.01       &  0.47$\pm$0.01      &  0.61$\pm$0.02      &  0.79$\pm$0.02     \\ 

\multicolumn{1}{l}{23} & \multicolumn{1}{l}{2E~1216+0700}       & \multicolumn{1}{l}{0.080}      &  6.45$\pm$0.17      &  7.79$\pm$0.18      &  17.42$\pm$0.46     &  24.85$\pm$1.65     &  -         &  - &  -        &  1.28$\pm$0.04       &  1.75$\pm$0.05      &  2.70$\pm$0.08      &  3.90$\pm$0.12     \\ 

\multicolumn{1}{l}{24} & \multicolumn{1}{l}{1RXS~J122019}       & \multicolumn{1}{l}{0.286}      &  2.08$\pm$0.06      &  3.06$\pm$0.08      &  5.66$\pm$0.20     &  7.83$\pm$1.09     &  0.52$\pm$0.06         &  0.78$\pm$0.10        &  1.35$\pm$0.11        &  -       &  -      &  -      &  -     \\ 

\multicolumn{1}{l}{25} & \multicolumn{1}{l}{LBQS~1228+1116}       & \multicolumn{1}{l}{0.236}      &  3.08$\pm$0.08      &  4.68$\pm$0.12      &  11.17$\pm$0.28     &  30.99$\pm$1.86     &  -         &  -        &  -        &  0.96$\pm$0.03       &  0.99$\pm$0.03      &  1.29$\pm$0.04      &  1.56$\pm$0.05     \\ 

\multicolumn{1}{l}{26} & \multicolumn{1}{l}{2XMM~J123126.4+105111}       & \multicolumn{1}{l}{0.304}      &  0.49$\pm$0.02      &  0.66$\pm$0.02      &  1.71$\pm$0.17     &  2.39$\pm$0.91     &  -         &  -        &  -        &  0.12$\pm$0.00       &  0.14$\pm$0.01      &  0.19$\pm$0.01      &  0.33$\pm$0.01     \\ 

\multicolumn{1}{l}{27} & \multicolumn{1}{l}{Mrk~0771}       & \multicolumn{1}{l}{0.064}      &  13.34$\pm$0.35      &  18.42$\pm$0.43      &  57.38$\pm$1.17     &  159.39$\pm$6.68     &  4.63$\pm$0.26         &  6.15$\pm$0.37        &  9.46$\pm$0.39        &  -       &  -      &  -      &  -     \\ 

\multicolumn{1}{l}{28} & \multicolumn{1}{l}{RX~J1233.9+0747}       & \multicolumn{1}{l}{0.371}      &  0.51$\pm$0.02      &  0.48$\pm$0.02      &  1.09$\pm$0.18     &  5.33$\pm$0.19     &  -         &  -        &  -        &  0.11$\pm$0.00       &  0.15$\pm$0.01      &  0.19$\pm$0.01      &  0.24$\pm$0.01     \\ 

\multicolumn{1}{l}{29} & \multicolumn{1}{l}{RX~J1236.0+2641}       & \multicolumn{1}{l}{0.209}      &  3.08$\pm$0.08      &  4.61$\pm$0.11      &  13.12$\pm$0.37     &  37.95$\pm$1.98     &  0.74$\pm$0.05         &  1.04$\pm$0.08        &  1.66$\pm$0.08        &  -       &  -      &  -      &  -     \\ 

\multicolumn{1}{l}{30} & \multicolumn{1}{l}{PG~1244+026}       & \multicolumn{1}{l}{0.048}      &  6.86$\pm$0.17      &  10.03$\pm$0.24      &  41.53$\pm$0.88     &  114.83$\pm$4.87     &  -         &  -        &  -        &  1.34$\pm$0.04       &  1.58$\pm$0.05      &  2.15$\pm$0.06      &  3.23$\pm$0.10     \\ 

\multicolumn{1}{l}{31} & \multicolumn{1}{l}{2XMM~J125553.0+272405}       & \multicolumn{1}{l}{0.316}      &  1.13$\pm$0.03      &  1.36$\pm$0.04      &  2.39$\pm$0.17     &  6.37$\pm$0.93     &  -         &  -        &  -        &  0.28$\pm$0.01       &  0.35$\pm$0.01      &  0.51$\pm$0.02      &  0.78$\pm$0.02     \\ 

\multicolumn{1}{l}{32} & \multicolumn{1}{l}{RBS 1201}       & \multicolumn{1}{l}{0.091}      &  3.20$\pm$0.08      &  3.59$\pm$0.09      &  8.29$\pm$0.25     &  19.65$\pm$1.43     &  -         &  -        &  -        &  1.03$\pm$0.03       &  1.27$\pm$0.04      &  1.74$\pm$0.05      &  1.92$\pm$0.06     \\ 

\multicolumn{1}{l}{33} & \multicolumn{1}{l}{2XMM~J132101.4+340658}       & \multicolumn{1}{l}{0.334}      &  0.77$\pm$0.02      &  0.94$\pm$0.03      &  1.73$\pm$0.11     &  3.93$\pm$0.78     &  0.31$\pm$0.05         &  0.29$\pm$0.06        &  0.61$\pm$0.08        &  -       &  -      &  -      &  -     \\ 

\multicolumn{1}{l}{34} & \multicolumn{1}{l}{1RXS~J132447}       & \multicolumn{1}{l}{0.306}      &  1.65$\pm$0.05      &  2.23$\pm$0.06      &  4.21$\pm$0.18     &  11.40$\pm$1.06     &  -         &  -        &  -        &  0.82$\pm$0.03       &  0.84$\pm$0.03      &  0.96$\pm$0.03      &  1.19$\pm$0.04     \\ 

\multicolumn{1}{l}{35} & \multicolumn{1}{l}{UM~602}       & \multicolumn{1}{l}{0.237}      &  3.74$\pm$0.10      &  5.19$\pm$0.13      &  8.34$\pm$0.23     &  18.25$\pm$1.31     &  -         &  -        &  -        &  0.52$\pm$0.02       &  0.57$\pm$0.02      &  0.80$\pm$0.02      &  1.47$\pm$0.04     \\ 

\multicolumn{1}{l}{36} & \multicolumn{1}{l}{1E~1346+26.7}       & \multicolumn{1}{l}{0.059}      &  4.34$\pm$0.13      &  4.87$\pm$0.13      &  18.74$\pm$0.42     &  34.30$\pm$1.86     &  -         &  -        &  -        &  1.09$\pm$0.03       &  1.40$\pm$0.04      &  2.12$\pm$0.06      &  2.66$\pm$0.08     \\ 

\multicolumn{1}{l}{37} & \multicolumn{1}{l}{PG~1352+183}       & \multicolumn{1}{l}{0.151}      &  7.09$\pm$0.18      &  10.07$\pm$0.24      &  18.03$\pm$0.41     &  35.78$\pm$1.80     &  1.61$\pm$0.08         &  2.11$\pm$0.10        &  3.78$\pm$0.14        &  -       &  -      &  -      &  -     \\ 

\multicolumn{1}{l}{38} & \multicolumn{1}{l}{Mrk~0464}       & \multicolumn{1}{l}{0.050}      &  3.69$\pm$0.10      &  5.03$\pm$0.12      &  14.32$\pm$0.34     &  28.24$\pm$1.59     &  1.94$\pm$0.08         &  2.33$\pm$0.10        &  3.04$\pm$0.12        &  -       &  -      &  -      &  -     \\ 

\multicolumn{1}{l}{39} & \multicolumn{1}{l}{1RXS~J135724}       & \multicolumn{1}{l}{0.106}      &  0.78$\pm$0.02      &  1.00$\pm$0.03      &  3.51$\pm$0.12     &  15.31$\pm$1.16     &  0.28$\pm$0.05         &  0.53$\pm$0.08      &  0.59$\pm$0.09        &  -       &  -      &  -      &  -     \\ 

\multicolumn{1}{l}{40} & \multicolumn{1}{l}{PG~1415+451}       & \multicolumn{1}{l}{0.114}      &  10.39$\pm$0.27      &  13.31$\pm$0.31      &  29.32$\pm$0.62     &  64.93$\pm$2.93     &  3.80$\pm$0.12         &  5.59$\pm$0.19        &  8.69$\pm$0.27        &  -       &  -      &  -      &  -     \\ 

\multicolumn{1}{l}{41} & \multicolumn{1}{l}{PG~1427+480}       & \multicolumn{1}{l}{0.221}      &  3.64$\pm$0.09      &  5.57$\pm$0.13      &  14.18$\pm$0.33     &  46.39$\pm$2.24     &  1.33$\pm$0.07         &  1.68$\pm$0.10        &  2.76$\pm$0.12        &  -       &  -      &  -      &  -     \\ 

\multicolumn{1}{l}{42} & \multicolumn{1}{l}{NGC~5683}       & \multicolumn{1}{l}{0.037}      &  5.37$\pm$0.14      &  4.96$\pm$0.12      &  19.55$\pm$0.45     &  50.03$\pm$2.32     &  5.46$\pm$0.27         &  6.47$\pm$0.37        &  7.10$\pm$0.29        &  -       &  -      &  -      &  -     \\ 

\multicolumn{1}{l}{43} & \multicolumn{1}{l}{RBS~1423}       & \multicolumn{1}{l}{0.208}      &  3.80$\pm$0.10      &  5.74$\pm$0.14      &  12.26$\pm$0.31     &  25.90$\pm$1.56     &  -         &  -        &  -        &  0.50$\pm$0.02       &  0.70$\pm$0.02      &  0.99$\pm$0.03      &  1.94$\pm$0.06     \\ 

\multicolumn{1}{l}{44} & \multicolumn{1}{l}{PG~1448+273}       & \multicolumn{1}{l}{0.065}      &  11.35$\pm$0.29      &  15.10$\pm$0.36      &  45.53$\pm$0.93     &  117.29$\pm$4.92     &  5.64$\pm$0.19         &  7.29$\pm$0.24        &  10.29$\pm$0.31        &  -       &  -      &  -      &  -     \\ 

\multicolumn{1}{l}{45} & \multicolumn{1}{l}{PG~1512+370}       & \multicolumn{1}{l}{0.371}      &  4.18$\pm$0.10      &  6.67$\pm$0.16      &  13.84$\pm$0.33     &  34.36$\pm$1.61     &  1.09$\pm$0.06         &  1.36$\pm$0.08        &  2.26$\pm$0.12        &  -       &  -      &  -      &  -     \\ 

\multicolumn{1}{l}{46} & \multicolumn{1}{l}{Q~1529+050}       & \multicolumn{1}{l}{0.218}      &  2.18$\pm$0.06      &  2.92$\pm$0.07      &  6.03$\pm$0.20     &  19.48$\pm$1.48     &  -         &  -        &  -        &  0.40$\pm$0.01       &  0.52$\pm$0.02      &  0.68$\pm$0.02      &  1.24$\pm$0.04     \\ 

\multicolumn{1}{l}{47} & \multicolumn{1}{l}{1E~1556+27.4}       & \multicolumn{1}{l}{0.090}      &  1.90$\pm$0.05      &  2.08$\pm$0.05      &  4.01$\pm$0.16     &  6.93$\pm$0.83     &  -         &  -        &  -        &  0.85$\pm$0.03       &  1.22$\pm$0.04      &  1.34$\pm$0.04      &  1.61$\pm$0.05     \\ 

\multicolumn{1}{l}{48} & \multicolumn{1}{l}{Mrk 0493}       & \multicolumn{1}{l}{0.031}      &  19.93$\pm$0.51      &  24.46$\pm$0.58      &  72.03$\pm$1.43     &  162.81$\pm$6.67     &  6.00$\pm$0.29         &  8.21$\pm$0.45        &  10.65$\pm$0.44        &  -       &  -      &  -      &  -     \\ 

\multicolumn{1}{l}{49} & \multicolumn{1}{l}{IIZw 177}       & \multicolumn{1}{l}{0.081}      &  3.79$\pm$0.10      &  4.72$\pm$0.11      &  20.70$\pm$0.48     &  60.99$\pm$2.76     &  1.77$\pm$0.09         &  2.56$\pm$0.13        &  2.91$\pm$0.15        &  -       &  -      &  -      &  -     \\ 

\multicolumn{1}{l}{50} & \multicolumn{1}{l}{PG~2233+134}       & \multicolumn{1}{l}{0.326}      &  5.25$\pm$0.13      &  8.26$\pm$0.20      &  20.97$\pm$0.50     &  57.39$\pm$2.92     &  0.99$\pm$0.06         &  1.18$\pm$0.07        &  2.03$\pm$0.12        &  -       &  -      &  -      &  -     \\ 

\multicolumn{1}{l}{51} & \multicolumn{1}{l}{Mrk~0926}       & \multicolumn{1}{l}{0.047}      &  69.60$\pm$1.80      &  86.16$\pm$2.03      &  123.69$\pm$2.53     &  256.37$\pm$10.75     &  8.97$\pm$0.36         &  12.43$\pm$0.49        &  17.00$\pm$0.64        &  -       &  -      &  -      &  -     \\ 

\hline
\hline
\end{tabular}
\end{center}
\end{table}
\end{landscape}

\section{Results}
\label{Sec4}

\subsection{Broadband SEDs \& Covering factors}

\begin{figure}
\begin{center}
\includegraphics[trim=0cm 0cm 0cm 0.0cm, clip=true, width =8.6cm, angle=0]
{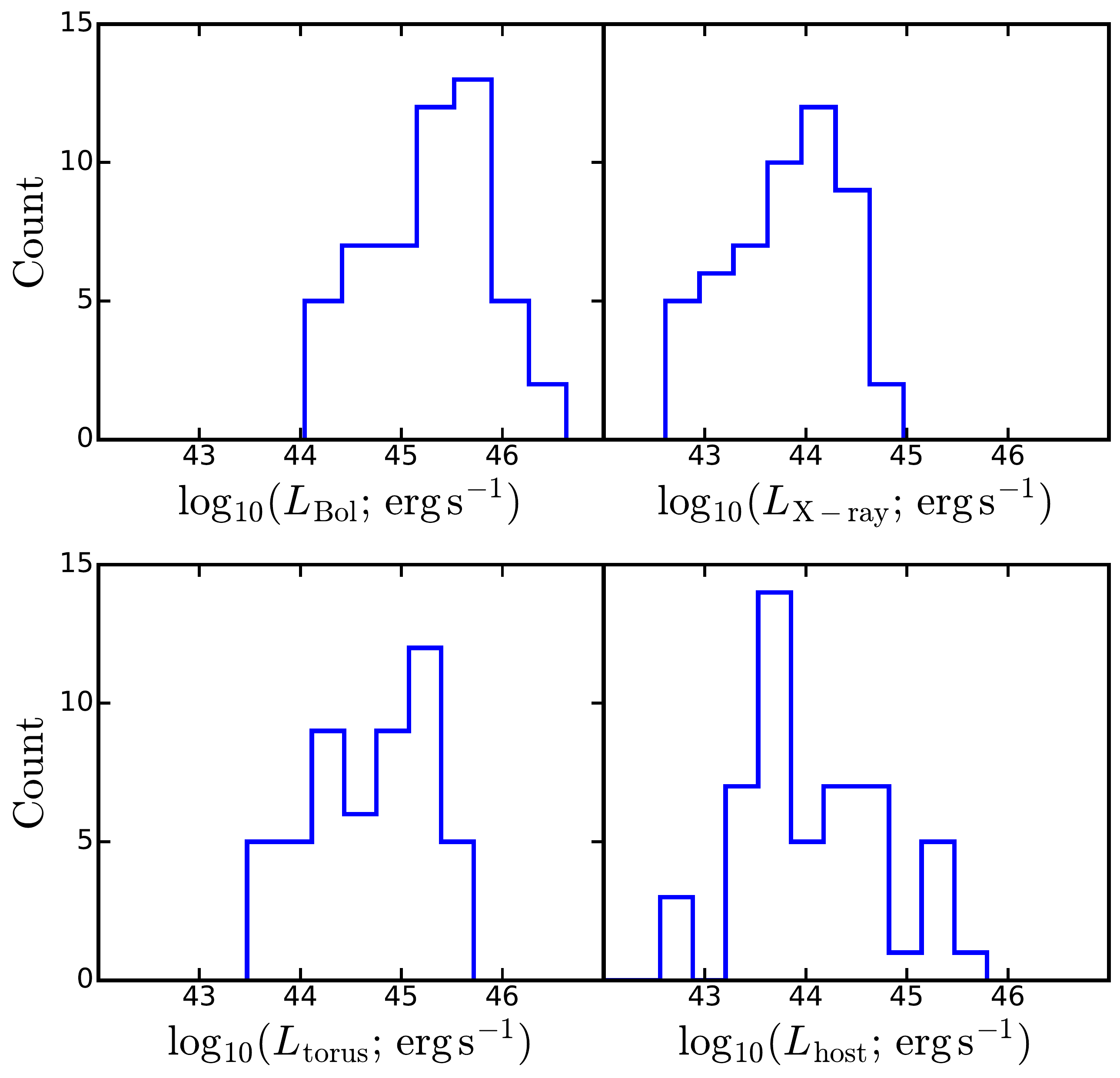}
\caption{\small The distribution of $L_{\rm Bol}$ (upper left), $L_{\rm X-ray}$ (2$-$10~keV) (upper right), $L_{\rm torus}$ (lower left) and $L_{\rm host}$ (lower right)}.
\label{Fig_lumin}
\end{center}
\end{figure}

Our best-fit model parameters for the SED fits are given in Table~\ref{tab_parameter}. We integrate the intrinsic {\sc optxagnf} model (after correcting for all absorption components) over the energy range of 10$^{-6}-$100~keV ($\sim$10$^{-5}-$1000~$\mu$m) to obtain $L_{\rm Bol}$, and over 2$-$10~keV to get the hard X-ray luminosity $L_{\rm X-ray}$. Now, in order to estimate the covering factor we can compare these to the IR luminosity of the torus $L_{\rm torus}$, which we find by integrating the \textit{agndust} model component over $\sim$1$-$1000~$\mu$m. We also give the luminosity of the host galaxy $L_{\rm host}$, obtained by integrating the host galaxy template over $\sim$0.1$-$1000~$\mu$m, and give its type in Table~\ref{tab_parameter}. The distributions of these luminosities are shown in Fig.~\ref{Fig_lumin}. The bolometric luminosities of most of the sources in our study are lower than the values of J12b by a factor of $\sim$1.7. This is due to a change in energy grid handling of {\sc optxagnf} in an older version of {\sc XSPEC}\footnote{See the {\sc XSPEC} patch 12.8.2j available at \url{https://heasarc.gsfc.nasa.gov/docs/xanadu/xspec/issues/archive/issues.12.8.2q.html}}. Though individual $L_{\rm Bol}$ values are affected, the overall sample trends remain the same. The host galaxy is significantly detected in 38 sources. We note that the host galaxy morphological types of 21 sources in our sample are known from the literature and many of these (No.~8, 13, 15, 16, 20, 21, 23, 27, 30, 37, 38, 39, 42, 44, 48, 50) are different to those indicated by our best-fit. We have refit these objects using the galaxy template fixed to the literature values. Although there is a marginal increase in $\chi^2$ when using the substituted templates, this has no significant impact on the parameters derived for either the AGN or the torus.

The primary goal of this work is to determine the distribution of covering factors $f_c$ of the unobscured AGN sample. This distribution of $f_c$ obtained from the ratio of the torus luminosity to the bolometric luminosity is given in Fig.~\ref{fc_hist}. The distribution, with an average value around 0.30, has a scatter of $\sim$0.17. The source with the minimum covering factor is 1E~1556+27.4 (No.~47). The value of $f_c \sim$0.02 suggests that this source could be a \textit{hot-dust-poor} AGN \citep{2010ApJ...724L..59H} characterised by weak IR emission. At the opposite extreme, 2E~1216+0700 (No.~23) has the maximum value of $f_c$ ($\sim 0.88$).

\subsection{Comparison with Previous Work}

Many previous studies have discussed the obscured AGN fraction. \cite{2010ApJ...714..561L} reviewed the studies that dealt with the fraction of obscured AGN and concluded that the obscured fraction for the non-X-ray selected samples has a mean value of about 0.58 with a dispersion of $\sim$~0.05. They also gave a rough estimate of 0.53 for the obscured fraction using the updated Swift/BAT hard X-ray catalogue \citep{2010ApJS..186..378T}, after applying the correction for missing Compton thick objects \citep{1999ApJ...522..157R}. The covering factors of a sample of 5281 \textit{WISE}, UKIDSS and SDSS selected high luminosity quasars ($L_{\rm Bol}>$10$^{46}$~erg~s$^{-1}$) with redshift~$<$~1.5 was determined by \cite{2013MNRAS.429.1494R}. They found that the covering factors (estimated by using the ratio of IR to UV/optical luminosity) obey a log-normal distribution with a mean observed value of $\sim$~0.39 and a dispersion of $\sim$~0.2. The study by \cite{2013ApJ...777...86L} estimated the covering factor for a sample of X-ray selected type~1 AGN with an even wider span of redshifts (0.10 $\leq z \leq 3.75$). They determined the covering factor of the optically-thin torus and this was observed to be in the range from $\sim0.45$ to $\sim0.75$. In comparison, the result from our sample is relatively low, especially since our sample is comprised only of Seyfert type~1 objects, so radiative transfer corrections, which reduce $f_c$ due to the torus being optically-thick to its own radiation at high inclination, should not be important (but see \citealt{2016MNRAS.458.2288S}). At high luminosities, two other studies, \cite{2011ApJ...737L..36M} and \cite{2011MNRAS.414..218L}, concentrated on the hot (NIR-emitting) dust component and found even lower hot dust covering factors of $\sim$0.13 and $\sim$0.07, respectively.

In APPENDIX~\ref{D}, we investigate the effect of varying the template SEDs on the $f_c$ measurements. In particular, we examined the impact of using two other IR SEDs for type~1 AGN $-$(1) proposed by \cite{2011MNRAS.414.1082M} based upon observations of local sources in which the AGN dominates the IR portion of the SED over the host galaxy contribution; and (2) a theoretical clumpy torus SED by \cite{2010A&A...523A..27H}. While the resultant mean $f_c$ value and the distribution of values from these template fits do not dramatically differ from those obtained previously when we used the SEDs from \cite{2004MNRAS.355..973S}, we found that both these sets of templates required an additional hot dust component in the NIR regime, suggesting that they are too restricted to account for the broadband features in our observations. This finding is interesting, but further investigation is beyond the scope of our present paper. However, we note that similar hot dust components are required in other quasar SED studies (e.g. \cite{2009ApJ...705..298M}).\\

\subsection{$f_c$ in Sub-samples}

\begin{figure}
\begin{center}
\includegraphics[trim=0cm 0cm 0cm 0.0cm, clip=true, width =8.6cm, angle=0]
{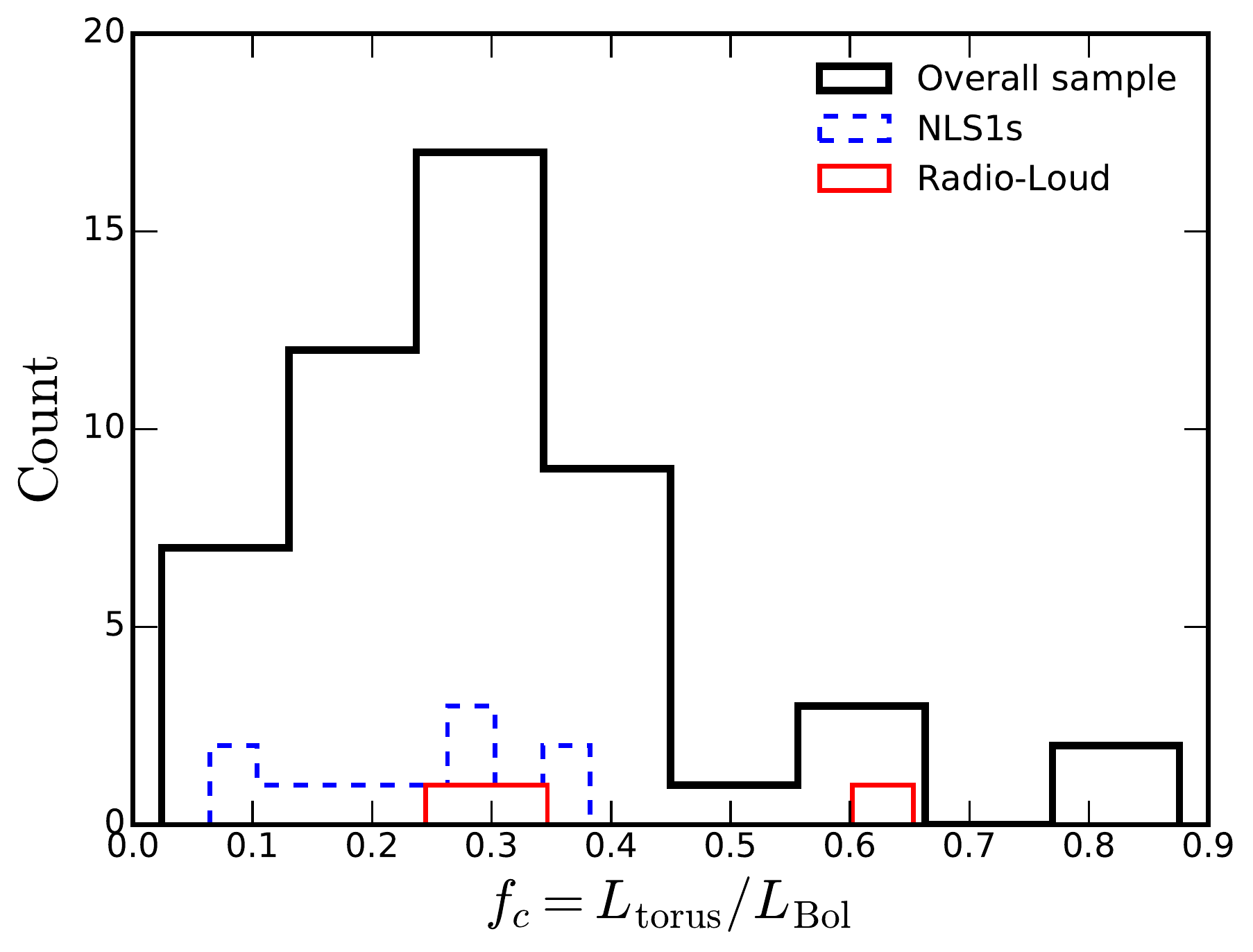}
\caption{\small Histogram of covering factors for the sample of 51 type~1 AGN (thick black line) and the sub-samples of 12 NLS1s (dashed blue line) and 3 radio-loud sources (thin red line).} 
\label{fc_hist}
\end{center}
\end{figure}

From our main sample, we have constructed sub-samples of three radio-loud objects and 12 NLS1 galaxies. The distributions of covering factor for these sub-samples are shown in Fig.~\ref{fc_hist}. 

\subsubsection{Radio-loud sources}

A significant contribution from a jet may reduce the covering factor in radio-loud sources. One of these objects in our sample, PG~1004+130 (No.~13), is a BAL quasar with a very weak X-ray spectrum. J12b suggests that the origin of X-rays in this source could be a sub-parsec-scale jet. However, the covering factors of the three radio-loud objects (No.~13, No.~14 \& No.~45) in our sample are 0.33, 0.65 and 0.24, respectively, indicating that the covering factors may not be affected by the radio-loudness/quietness of the sources, although we cannot draw general conclusions from such a small sample of radio-loud sources.

\subsubsection{NLS1 galaxies}

It is known that the NLS1s, in general, tend to have higher values of Eddington ratio. Since there is an anti-correlation between $f_c$ and $\lambda_{\rm Edd}$ we expect the 12 NLS1s in our sample to have low values for $f_c$. The covering factors of NLS1s in our sample range from $\sim$0.06 to $\sim$0.38 with a mean value around 0.23 and a dispersion of about 0.1. The Kolmogorov-Smirnov test (K-S test) with a probability of $\sim$75.5\% reveals that the distribution of $f_c$ in NLS1s does not differ significantly from that of the overall sample. But we caution that no strong statement can be made, based on such a small sub-sample of NLS1s. We need further investigations to get better constraints on this result.

\subsection{Correlations obtained}

\begin{figure}
\begin{center}
\includegraphics[trim=0cm 0cm 0cm 0.0cm, clip=true, width =9.0cm, angle=0]{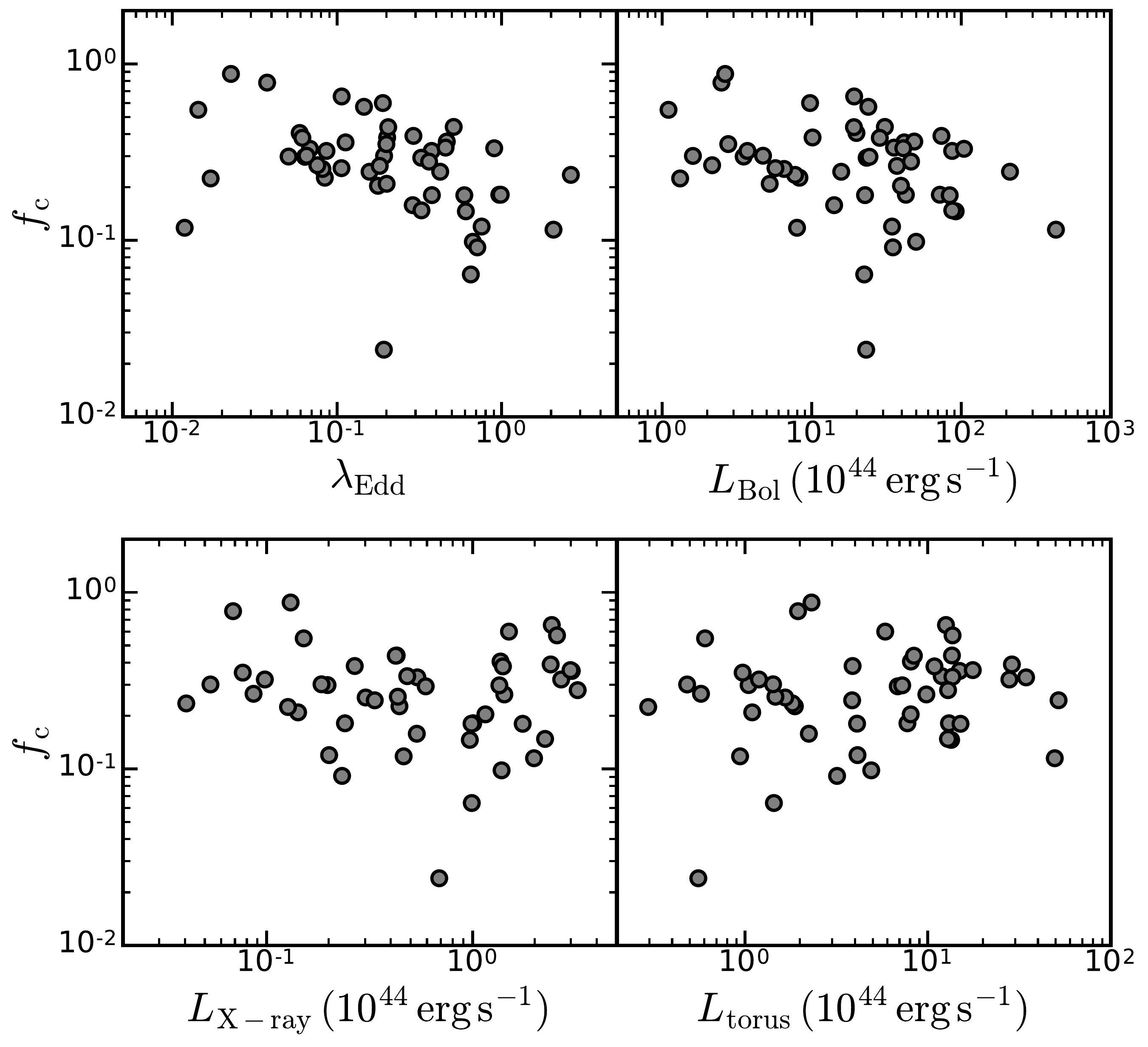}
\caption{\small The variation of $f_c$ with $\lambda_{\rm Edd}$ (upper left), $L_{\rm Bol}$ (upper right), $L_{\rm X-ray}$ (lower left) and $L_{\rm torus}$ (lower right).}
\label{Fig_fc_corr}
\end{center}
\end{figure}

J12b carried out a systematic study of the correlation between the different AGN parameters in this sample. We follow J12b, but also include the new torus parameters, $L_{\rm torus}$ and covering fraction $f_{c}$ and compute the correlations using the Spearman's rank-order method \citep{numerical_recipes}. The rank coefficient $\rho_s$ and probability $d_s$ (also known as the p-value) for the Spearman's correlation between different parameters are listed in Table~\ref{tab_corr}. We recover most of the correlations, obtained by J12b, between $L_{\rm Bol}$, $\lambda_{\rm Edd}$, $\Gamma$ and the hard X-ray bolometric correction $\kappa_{2-10}$ defined as $L_{\rm Bol}$/$L_{\rm X-ray}$ \citep{2007MNRAS.381.1235V}. Additionally, we find a marginal correlation of $f_c$ with $L_{\rm Bol}$ ($d_s$=0.05) and a stronger correlation between $f_c$ and $\lambda_{\rm Edd}$ ($d_s$=0.002). But $f_c$ shows no significant correlation with $L_{\rm X-ray}$ or $L_{\rm torus}$, as shown in Fig.~\ref{Fig_fc_corr}. 

J12b suggest that the combination of $\lambda_{\rm Edd}$ and M$_{\rm BH}$ drives the correlations seen in the AGN parameters, with $\lambda_{\rm Edd}$ changing the accretion flow geometry (resulting in the correlations in $\Gamma$, $R_{\rm cor}$ and $f_{\rm pl}$ which drive the correlation with $\kappa_{2-10}$) while M$_{\rm BH}$ and $\lambda_{\rm Edd}$ together set the overall luminosity and 
the peak temperature of the disc. As $f_c$ correlates with $\lambda_{\rm Edd}$ and $L_{\rm Bol}\propto$~M$_{\rm BH}\lambda_{\rm Edd}$ then it is clear that the correlations of $f_c$ with $\lambda_{\rm Edd}$ and $L_{\rm Bol}$, shown in Fig.~\ref{Fig_fc_corr}, will also give rise to correlations of $f_c$ with the other AGN parameters. However, the major statistical correlation with the new torus parameters is that $f_c$ correlates with $\lambda_{\rm Edd}$ and $L_{\rm Bol}$. This is in accord with the physical expectations that $\lambda_{\rm Edd}$ and $L_{\rm Bol}$ are the two key parameters which determine the properties of the accretion flow. 

\begin{figure}
\begin{center}
\includegraphics[trim=0cm 0cm 0cm 0.0cm, clip=true, width =8.8cm, angle=0]
{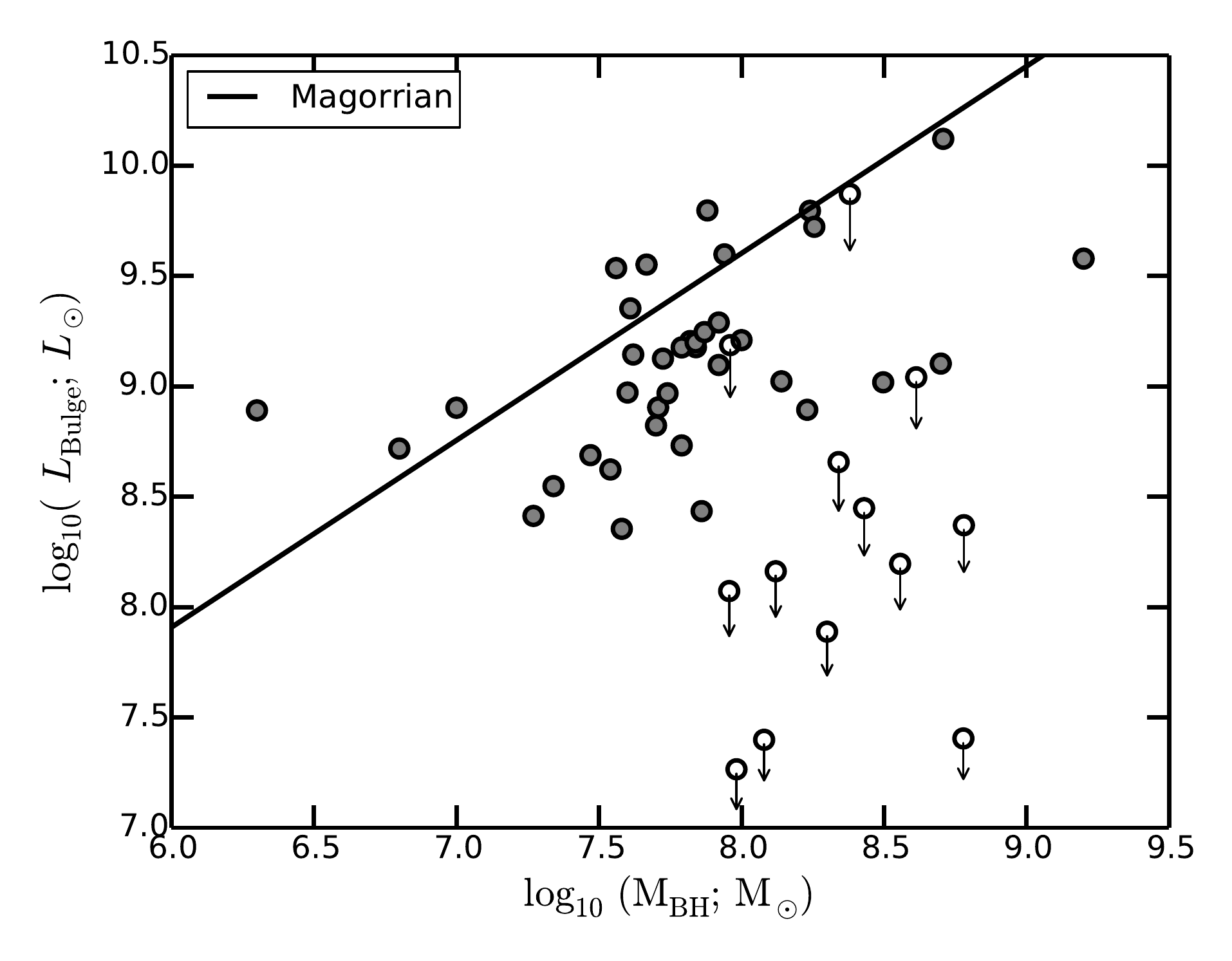}
\caption{\small Plot showing the relationship between the bulge luminosity and the black hole mass. The circles are the data points and the solid black line describes the Margorrian relationship \citep{1998AJ....115.2285M} for our data. The open circles with lower arrow denote the sources for which we have considered the upper limits of $L_{\rm host}$.}
\label{Fig_magorrian}
\end{center}
\end{figure}

There is also a weak correlation between $L_{\rm host}$ and M$_{\rm BH}$. This is expected from the black hole mass and bulge luminosity ($L_{\rm Bulge}$) 
relation of \cite{1998AJ....115.2285M}. Since the SED of a normal galaxy stellar population peaks in the H-band, we calculated the corresponding luminosity for our sample in terms of $L_{\rm host}$ and the relative H-band flux from \cite{2007ApJ...663...81P} SED templates. Here, we are not considering Mrk~0110 (No.~9) since $L_{\rm host}$ is not well constrained for this source. We find that the sources in which our SED fits include a significant host galaxy contribution approximately follow the Magorrian relationship, though there is significant scatter ($\sim$0.5~dex) between M$_{\rm BH}$ and $L_{\rm Bulge}$ (see Fig.~\ref{Fig_magorrian}). This shows that our host galaxy fitted components are largely reasonable. Here, we note that among the 13 sources for which there is no significant host galaxy contribution, most of them are fitted by starburst galaxy templates (e.g. No.~50, No.~16). These starburst templates peak in the FIR and show a minimum around NIR wavelengths. Since our SEDs do not cover the FIR wavelengths and the starbursts tend to be extended we may be underestimating their FIR emission, and also their bulge luminosities. In these few cases, we cannot rule out that we are then over-estimating the torus contribution and hence also their $f_c$ values. When excluding the 13 sources for which $L_{\rm host}$ are not constrained, neither the scatter ($\sim0.17$) nor the mean value ($\sim0.33$) for the remaining 38 sources are changed significantly from the covering factors of the overall sample. This indicates that the uncertainties in $L_{\rm host}$ do not strongly bias our overall results.

\begin{landscape}
\begin{table}
\scriptsize
\caption{\scriptsize Broadband SED fitting results. \textit{hostpol} model: best-fit \textit{hostpol} template ($^*$starburst galaxies); $\chi^2_{red}$: reduced $\chi^2$ for the best-fit model; N$_{\rm H}^{\rm Gal}~\&$ $N_{\rm H}^{\rm Int}$: Galactic and intrinsic column densities, respectively, in 10$^{20}$cm$^{-2}$; M$_{\rm BH}$: black hole mass in 10$^7$M$_\odot$ (fixed to the best-fit values of J12b); $\lambda_{\rm Edd}$: Eddington ratio; R$_{\rm cor}$: coronal radius in R$_g$; kT$_e$: electron temperature for the soft Comptonisation component in keV; $\tau$: optical depth of the soft Comptonisation component; $\Gamma$: hard X-ray photon index; f$_{\rm pl}$: fraction of the power below R$_{\rm cor}$ emitted in the hard Comptonisation component; $L_{\rm X-ray}$: unabsorbed X-ray luminosity in the band of 2$-$10~keV; $L_{\rm host}$: host galaxy luminosity in the $\sim$0.1$-$1000~$\mu$m band; $L_{\rm torus}$: infrared ($\sim$1$-$1000~$\mu$m) luminosity of the torus emission; $L_{\rm Bol}$: bolometric luminosity in the range of 10$^{-6}-$100~keV; $\kappa_{2-10}$: hard X$-$ray bolometric correction ($L_{\rm Bol}/L_{\rm X-ray}$); $f_c$: Covering factor, the ratio of $L_{\rm torus}$ and $L_{\rm Bol}$. All luminosities are expressed in 10$^{44}$ erg s$^{-1}$.}
\begin{center}
\begin{tabular}{rlcccccccccccccccccc}
\hline
\hline
\\
\multicolumn{1}{c}{No.} & \multicolumn{1}{c}{\textit{hostpol}} & \multicolumn{1}{c}{$\chi_{red}^2$} & \multicolumn{1}{c}{$N_{\rm H}^{\rm Gal}$} & \multicolumn{1}{c}{$N_{\rm H}^{\rm Int}$} & \multicolumn{1}{c}{M$_{\rm BH}$} & \multicolumn{1}{c}{$\lambda_{\rm Edd}$} & \multicolumn{1}{c}{R$_{\rm cor}$} & \multicolumn{1}{c}{kT$_e$} & \multicolumn{1}{c}{$\tau$} & \multicolumn{1}{c}{$\Gamma$} & \multicolumn{1}{c}{f$_{\rm pl}$}  & \multicolumn{1}{c}{$L_{\rm X-ray}$} & \multicolumn{1}{c}{$L_{\rm host}$} & \multicolumn{1}{c}{$L_{\rm torus}$} & \multicolumn{1}{c}{$L_{\rm Bol}$} & \multicolumn{1}{c}{$\kappa_{2-10}$} & \multicolumn{1}{c}{$f_c$} \\
\\
\multicolumn{1}{c}{} & \multicolumn{1}{c}{model} & \multicolumn{1}{c}{} & \multicolumn{1}{c}{10$^{20}$cm$^{-2}$} & \multicolumn{1}{c}{10$^{20}$cm$^{-2}$} & \multicolumn{1}{c}{10$^7$ M$_{\odot}$} & \multicolumn{1}{c}{} & \multicolumn{1}{c}{R$_g$} & \multicolumn{1}{c}{keV} & \multicolumn{1}{c}{} & \multicolumn{1}{c}{} & \multicolumn{1}{c}{} & \multicolumn{1}{c}{10$^{44}$ erg s$^{-1}$} & \multicolumn{1}{c}{10$^{44}$ erg s$^{-1}$} & \multicolumn{1}{c}{10$^{44}$ erg s$^{-1}$} & \multicolumn{1}{c}{10$^{44}$ erg s$^{-1}$} & \multicolumn{1}{c}{} & \multicolumn{1}{c}{} \\
\hline
\\
1 & E5 & 1.34 & 1.79 & 3.43 & 41.00 & 0.11 & 97.0 & 0.11 & 28.1 & 1.85 & 0.49 & 3.0 & $<$0.5 & 14.9 & 41.6 & 13.74 & 0.36 \\
2 & S0 & 1.99 & 2.43 & 1.29 & 6.92 & 0.08 & 94.2 & 0.19 & 16.9 & 1.82 & 0.36 & 0.4 & 0.6 & 1.9 & 8.3 & 18.80 & 0.23 \\
3 & S0 & 1.34 & 6.31 & 0.66 & 4.07 & 0.20 & 12.6 & 0.18 & 15.2 & 1.87 & 0.82 & 0.3 & 0.9 & 3.9 & 10.1 & 37.94 & 0.38 \\
4 & IRAS~22491-1808$^*$ & 1.23 & 3.49 & 2.69 & 60.00 & 0.01 & 84.3 & 0.34 & 14.3 & 1.67 & 0.45 & 0.5 & $<$0.4 & 0.9 & 8.0 & 17.28 & 0.12 \\
5 & S0 & 1.29 & 3.53 & 8.08 & 8.71 & 0.97 & 34.8 & 0.20 & 14.3 & 2.16 & 0.15 & 1.0 & 1.6 & 13.1 & 72.0 & 71.60 & 0.18 \\
6 & IRAS~22491-1808$^*$ & 1.39 & 4.24 & 0.00 & 31.40 & 0.29 & 21.0 & 0.42 & 11.0 & 1.85 & 0.45 & 2.4 & 17.1 & 28.8 & 73.8 & 30.90 & 0.40 \\
7 & IRAS~22491-1808$^*$ & 1.20 & 1.33 & 0.00 & 3.80 & 0.99 & 10.5 & 0.15 & 32.7 & 2.20 & 0.31 & 0.2 & 3.7 & 7.7 & 42.7 & 178.10 & 0.18 \\
8 & S0 & 1.89 & 3.12 & 3.80 & 3.47 & 0.29 & 21.7 & 0.18 & 14.4 & 1.80 & 0.51 & 0.5 & 0.2 & 2.2 & 14.1 & 26.42 & 0.16 \\
9 & Sa & 19.62 & 1.30 & 0.00 & 2.51 & 0.65 & 17.4 & 0.26 & 15.5 & 1.73 & 0.83 & 1.0 & $-$ & 1.4 & 22.5 & 22.70 & 0.06 \\
10 & S0 & 1.70 & 1.74 & 1.01 & 21.90 & 0.38 & 39.5 & 0.49 & 9.6 & 1.76 & 0.28 & 2.7 & $<$0.2 & 27.9 & 87.0 & 32.33 & 0.32 \\
11 & IRAS 20551-4250$^*$ & 1.00 & 1.72 & 0.00 & 8.32 & 0.38 & 23.1 & 0.45 & 12.3 & 1.54 & 0.84 & 1.0 & 8.1 & 4.1 & 22.8 & 23.05 & 0.18 \\
12 & S0 & 1.17 & 1.20 & 0.00 & 6.61 & 0.05 & 19.3 & 0.10 & 54.1 & 1.94 & 0.88 & 0.2 & 0.7 & 1.1 & 3.5 & 17.81 & 0.30 \\
13 & IRAS~22491-1808$^*$ & 1.92 & 3.56 & 0.00 & 158.49 & 0.07 & 9.9 & 0.19 & 15.0 & 1.45 & 0.93 & 0.5 & 62.1 & 34.5 & 104.5 & 193.86 & 0.33 \\
14 & S0 & 1.47 & 1.76 & 0.00 & 17.38 & 0.11 & 100.0 & 0.23 & 20.3 & 1.77 & 0.81 & 2.4 & 2.5 & 12.6 & 19.3 & 7.95 & 0.65 \\
15 & Sd & 1.67 & 1.31 & 2.81 & 0.20 & 2.64 & 100.0 & 0.30 & 8.7 & 2.20 & 0.03 & 0.04 & 0.8 & 1.8 & 7.8 & 190.64 & 0.23 \\
16 & IRAS~22491-1808$^*$ & 1.50 & 1.70 & 2.21 & 19.95 & 0.18 & 28.9 & 0.30 & 13.0 & 1.74 & 0.34 & 1.2 & $<$1.3 & 8.1 & 39.6 & 34.33 & 0.20 \\
17 & Arp~220$^*$ & 1.06 & 0.65 & 1.13 & 6.17 & 0.51 & 93.8 & 0.48 & 9.0 & 1.61 & 0.12 & 0.4 & 18.0 & 13.6 & 30.9 & 72.26 & 0.44 \\
18 & E5 & 1.82 & 1.45 & 0.00 & 13.18 & 0.61 & 11.6 & 0.25 & 14.0 & 2.20 & 0.47 & 1.0 & $<$0.1 & 13.4 & 91.9 & 94.83 & 0.15 \\
19 & S0 & 1.00 & 3.70 & 0.00 & 5.10 & 0.04 & 69.1 & 0.16 & 20.6 & 1.94 & 0.21 & 0.1 & 0.3 & 2.0 & 2.5 & 36.39 & 0.78 \\
20 & Sd & 1.24 & 1.91 & 0.00 & 0.63 & 0.19 & 18.0 & 0.20 & 22.5 & 2.20 & 0.57 & 0.1 & 0.5 & 0.5 & 1.6 & 30.18 & 0.30 \\
21 & M~82$^*$ & 3.82 & 1.77 & 0.00 & 9.60 & 0.46 & 23.7 & 0.21 & 18.6 & 1.77 & 0.78 & 3.0 & $<$0.1 & 17.6 & 48.6 & 16.27 & 0.36 \\
22 & S0 & 1.21 & 2.75 & 4.37 & 6.90 & 0.08 & 77.7 & 0.12 & 23.1 & 1.82 & 0.33 & 0.3 & 0.6 & 1.7 & 6.6 & 21.68 & 0.25 \\
23 & S0 & 2.10 & 1.59 & 0.00 & 10.00 & 0.02 & 100.0 & 0.36 & 14.1 & 1.52 & 0.49 & 0.1 & 0.7 & 2.3 & 2.6 & 20.23 & 0.88 \\
24 & S0 & 2.90 & 1.63 & 0.00 & 18.00 & 0.15 & 39.8 & 0.22 & 25.1 & 1.91 & 0.95 & 2.6 & 2.1 & 13.7 & 23.9 & 9.32 & 0.57 \\
25 & IRAS~22491-1808$^*$ & 2.00 & 2.34 & 0.01 & 26.92 & 0.33 & 16.2 & 0.34 & 13.0 & 1.79 & 0.52 & 2.3 & $<$4.6 & 12.9 & 87.2 & 38.79 & 0.15 \\
26 & S0 & 1.20 & 2.31 & 6.50 & 5.01 & 0.76 & 16.5 & 0.35 & 9.2 & 2.01 & 0.1 & 0.2 & 0.3 & 4.1 & 34.5 & 171.75 & 0.12 \\
27 & IRAS~22491-1808$^*$ & 1.68 & 2.75 & 1.59 & 7.24 & 0.16 & 28.3 & 0.11 & 20.1 & 2.20 & 0.29 & 0.3 & 4.5 & 3.9 & 15.8 & 47.15 & 0.24 \\
28 & S0 & 1.64 & 1.45 & 5.55 & 9.12 & 0.67 & 75.3 & 0.37 & 10.6 & 1.69 & 0.22 & 1.4 & 0.6 & 4.9 & 50.0 & 36.17 & 0.10 \\
29 & Arp~220$^*$ & 1.23 & 1.18 & 1.31 & 7.41 & 0.46 & 12.9 & 0.15 & 15.0 & 2.18 & 0.45 & 0.5 & 21.2 & 11.9 & 35.6 & 73.93 & 0.34 \\
30 & IRAS~22491-1808$^*$ & 1.24 & 1.87 & 0.00 & 1.86 & 0.20 & 19.0 & 0.23 & 19.3 & 2.20 & 0.46 & 0.1 & 4.2 & 1.1 & 5.3 & 36.96 & 0.21 \\
31 & Sb & 1.55 & 0.84 & 0.00 & 50.00 & 0.06 & 100.0 & 0.40 & 14.2 & 1.45 & 0.74 & 1.4 & $<$0.7 & 8.1 & 19.9 & 14.59 & 0.41 \\
32 & S0 & 1.64 & 0.90 & 0.00 & 4.17 & 0.11 & 100.0 & 0.34 & 13.3 & 1.85 & 0.49 & 0.4 & 0.6 & 1.5 & 5.7 & 13.23 & 0.26 \\
33 & S0 & 2.05 & 1.07 & 0.00 & 8.32 & 0.33 & 13.6 & 0.20 & 20.2 & 2.19 & 0.79 & 0.6 & 0.8 & 6.9 & 23.3 & 39.45 & 0.29 \\
34 & S0 & 1.48 & 1.83 & 1.96 & 51.00 & 0.06 & 54.7 & 0.23 & 16.4 & 1.87 & 0.40 & 1.4 & 5.3 & 10.9 & 28.6 & 20.36 & 0.38 \\
35 & S0 & 2.67 & 1.76 & 0.00 & 4.64 & 0.90 & 77.5 & 0.31 & 18.4 & 1.90 & 0.97 & 6.2 & 1.4 & 13.6 & 41.0 & 6.67 & 0.33 \\
36 & Sb & 1.43 & 1.18 & 2.48 & 1.00 & 0.20 & 25.6 & 0.44 & 7.9 & 2.19 & 0.34 & 1.0 & 0.4 & 1.0 & 2.8 & 36.25 & 0.35 \\
37 & Sa & 1.84 & 1.82 & 0.98 & 17.00 & 0.18 & 93.1 & 0.16 & 20.1 & 2.12 & 0.39 & 1.4 & 0.3 & 9.8 & 37.2 & 26.10 & 0.26 \\
38 & S0 & 1.56 & 1.42 & 0.65 & 6.17 & 0.01 & 100.0 & 0.23 & 14.1 & 1.72 & 0.90 & 0.2 & 0.2 & 0.6 & 1.1 & 7.31 & 0.55 \\
39 & Arp~220$^*$ & 1.01 & 1.36 & 1.45 & 2.19 & 0.08 & 91.9 & 0.22 & 17.5 & 2.08 & 0.35 & 0.1 & 4.2 & 0.6 & 2.2 & 25.06 & 0.27 \\
40 & S0 & 2.15 & 0.77 & 0.00 & 7.59 & 0.20 & 15.5 & 0.35 & 11.7 & 1.93 & 0.45 & 0.4 & 2.5 & 8.4 & 19.2 & 45.35 & 0.44 \\
41 & IRAS~22491-1808$^*$ & 1.48 & 1.81 & 0.82 & 13.80 & 0.59 & 14.0 & 0.39 & 11.0 & 1.86 & 0.51 & 1.8 & 17.3 & 15.1 & 83.8 & 47.91 & 0.18 \\
42 & Sc & 2.56 & 2.86 & 0.00 & 5.50 & 0.02 & 47.9 & 0.24 & 15.0 & 1.89 & 0.79 & 0.1 & 0.6 & 0.3 & 1.3 & 10.42 & 0.22 \\
43 & IRAS~22491-1808$^*$ & 1.54 & 2.69 & 0.23 & 12.00 & 0.36 & 100.0 & 0.29 & 13.3 & 1.87 & 0.46 & 3.2 & $<$0.4 & 12.9 & 46.2 & 14.28 & 0.28 \\
44 & E2 & 1.44 & 2.78 & 4.11 & 3.63 & 0.71 & 12.7 & 0.33 & 10.0 & 2.14 & 0.2 & 0.2 & 1.7 & 3.2 & 35.0 & 150.67 & 0.09 \\
45 & IRAS 20551-4250$^*$ & 2.54 & 1.46 & 2.26 & 60.26 & 0.42 & 49.7 & 0.20 & 16.9 & 1.93 & 0.37 & 9.3 & $<$1.5 & 51.9 & 212.0 & 22.85 & 0.24 \\
46 & IRAS~22491-1808$^*$ & 1.59 & 4.02 & 11.36 & 36.00 & 0.06 & 100.0 & 0.11 & 31.5 & 1.93 & 0.40 & 1.4 & $<$2.6 & 7.2 & 24.4 & 18.07 & 0.30 \\
47 & E2 & 1.41 & 3.78 & 17.63 & 9.05 & 0.19 & 100.0 & 0.11 & 29.7 & 1.87 & 0.20 & 0.7 & $<$0.1 & 0.6 & 23.2 & 33.66 & 0.02 \\
48 & S0 & 1.42 & 2.11 & 0.00 & 2.95 & 0.09 & 35.1 & 0.37 & 11.6 & 1.91 & 0.24 & 0.1 & 0.2 & 1.2 & 3.7 & 38.05 & 0.32 \\
49 & M~82$^*$ & 1.51 & 4.90 & 0.00 & 5.30 & 0.07 & 100.0 & 0.19 & 24.4 & 2.20 & 0.50 & 0.2 & 3.7 & 1.4 & 4.7 & 25.65 & 0.30 \\
50 & IRAS~22491-1808$^*$ & 5.17 & 4.51 & 0.00 & 23.99 & 2.07 & 9.3 & 0.62 & 7.3 & 2.20 & 0.46 & 2.0 & $<$20.5 & 49.4 & 430.3 & 216.60 & 0.11 \\
51 & S0 & 2.25 & 2.91 & 0.87 & 3.98 & 0.19 & 100.0 & 0.13 & 39.6 & 1.78 & 0.96 & 1.5 & 0.4 & 5.9 & 9.8 & 6.51 & 0.60 \\
\hline
\hline
\label{tab_parameter}
\end{tabular}
\end{center}
\end{table}
\end{landscape}

\begin{figure}
\begin{center}
\includegraphics[trim=0cm 0cm 0cm 0.0cm, clip=true, width =8.5cm, angle=0]{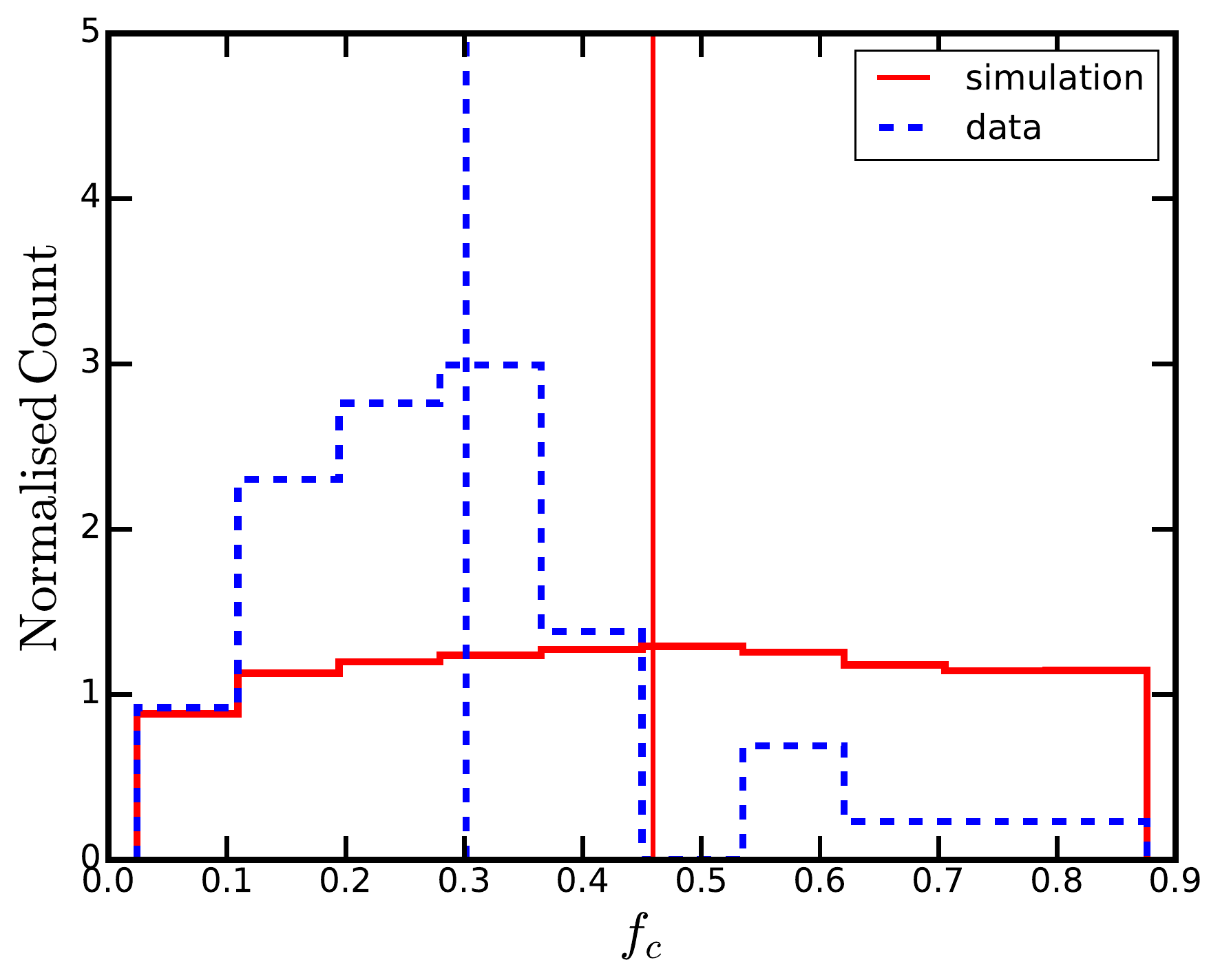}
\caption{\small Histogram of $f_c$ for the original sample (blue dashed line) and the simulation (red solid line). Blue (dashed) and red (solid) vertical lines show the mean values of the respective distributions of $f_c$.}
\label{simulation_fc}
\end{center}
\end{figure}

\begin{figure}
\begin{center}
\includegraphics[trim=0cm 0cm 0cm 0.0cm, clip=true, width =8.5cm, angle=0]{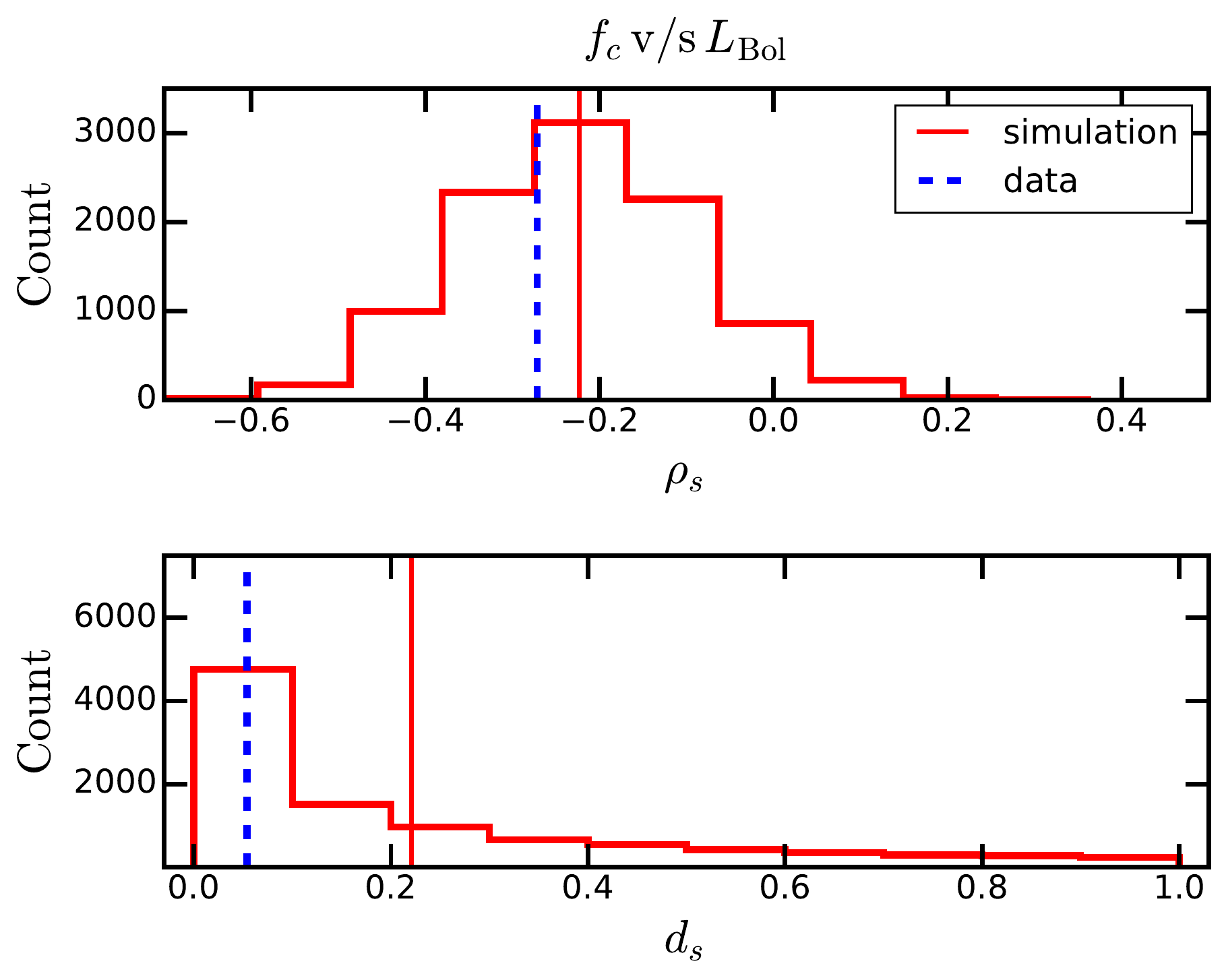}
\caption{\small Histogram of $\rho_s$ (upper panel) and $d_s$ (lower panel) of Spearman's correlation between $f_c$ and $L_{\rm Bol}$ for the simulation. The blue dashed line denotes $\rho_s$ of the original data (upper panel), and $\sim$36\% of the realisations lie below this showing stronger anti-correlation than the original sample.}
\label{simulation_fcLbol}
\end{center}
\end{figure}

\begin{figure}
\begin{center}
\includegraphics[trim=0cm 0cm 0cm 0.0cm, clip=true, width =8.6cm, angle=0]
{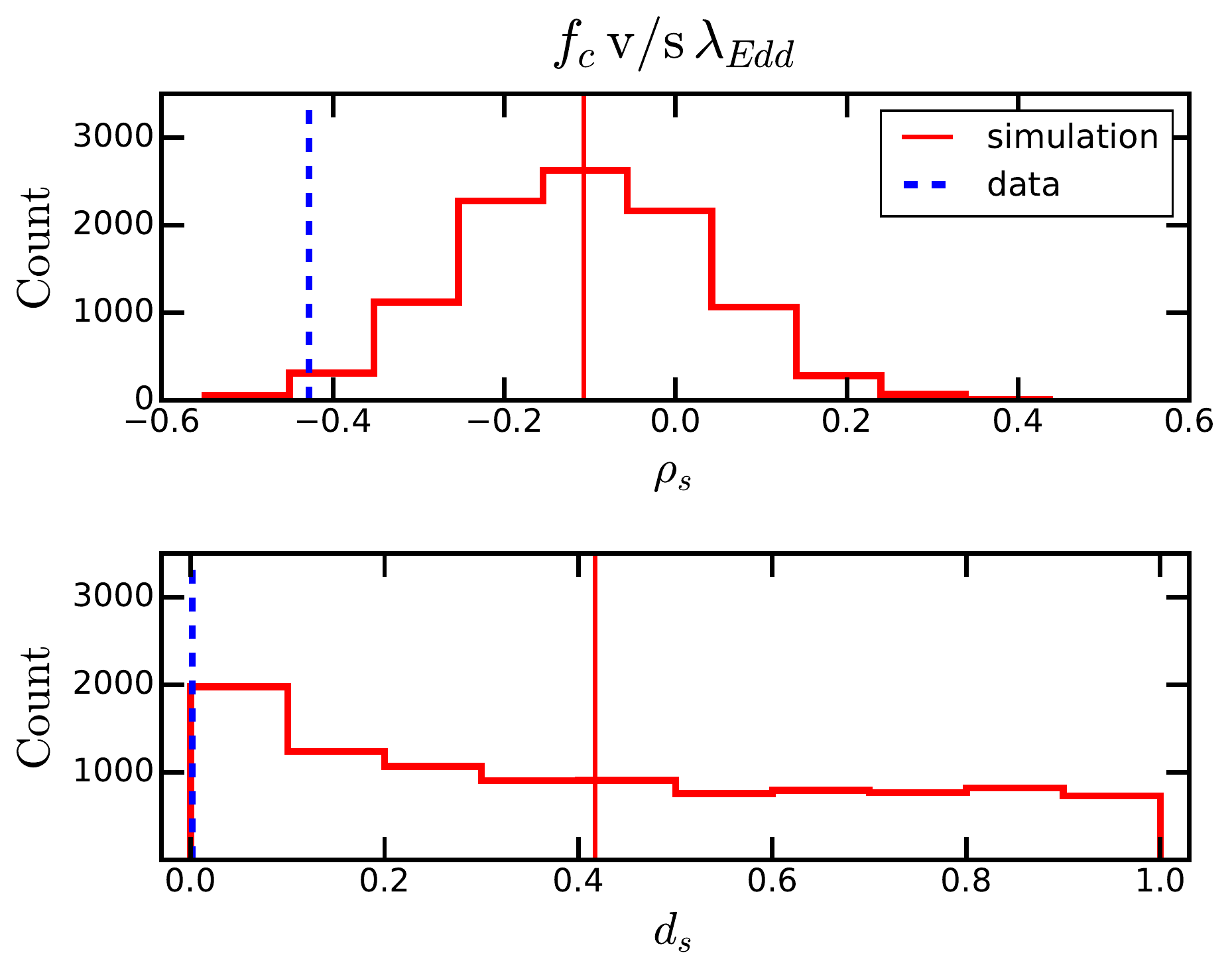}
\caption{\small Histogram of $\rho_s$ (upper panel) and $d_s$ (lower panel) of Spearman's correlation between $f_c$ and $\lambda_{\rm Edd}$ for 10000 realisations. Here, only $\sim$1\% of the realisations show stronger anti-correlation than the original data.}
\label{simulation_fcmdot}
\end{center}
\end{figure}

\section{Discussion}
\label{Sec5}

The effect of the well-known \textit{receding torus model} \citep{1991MNRAS.252..586L} 
predicts that $f_c$ decreases with increasing $L_{\rm Bol}$. Whilst our data marginally supports this prediction we find a stronger anti-correlation between $f_c$ and $\lambda_{\rm Edd}$. So it is not clear which parameter (or both) is the fundamental driver of the trends seen. This is made more complex as $f_c=L_{\rm torus}/L_{\rm Bol}$, so there is an implicit bias where $f_c$ will anti-correlate with $L_{\rm Bol}$.

In order to investigate this, we carried out a simulation to predict the correlation between these parameters given our sample. We generated 10000 realisations of random data where pairs of $L_{\rm Bol}$ and $\lambda_{\rm Edd}$ are drawn from the data using bootstrap sampling with replacement and then assigned a redshift from the sample, $z$, again with replacement. A sample of 51 $f_c$ values is produced from a uniform distribution within the range of $0.0-0.9$, roughly corresponding to the observational range. This allows us to calculate the IR flux which would result from assuming a random distribution of covering fraction. We applied an effective IR detection threshold to the simulations, similar to what is expected for the data. We use the standard 12~$\mu$m limiting flux of 1~mJy for the \textit{WISE} all-sky survey as our threshold \citep{2010AJ....140.1868W}. If the simulated IR flux is more than 1~mJy and the corresponding IR luminosity is within the range of the original data, the source is retained; otherwise, it is discarded and new values of $L_{\rm Bol}$, $\lambda_{\rm Edd}$, $z$ and $f_c$ are drawn. We checked that this gives rise to realisations with the same distribution of $L_{\rm torus}$ as found in the actual data. There is a difference at the level of 0.02 according to the K-S test, which is only marginally significant. Further tests carried out with randomised log($f_c$), make the difference even less significant. However, Fig.~\ref{simulation_fc} shows that the distribution of $f_c$ in the data is very different from the simulation, where an assumed initial random distribution is modified by the IR selection.

In each realisation, we calculate the Spearman's correlation between $L_{\rm Bol}$ and $f_c$. The rank coefficient of the original data was $-0.27$, very similar to the value of $\rho_s \sim -0.29$ seen in the simulation, and $\sim$36\% of the realisations show an anti-correlation stronger than that found in the actual data (see Fig.~\ref{simulation_fcLbol}). So, in our sample, there is no significant anti-correlation between $L_{\rm Bol}$ and $f_c$. 

We repeat this procedure for $\lambda_{\rm Edd}$ and $f_c$, but here the results are very different. The rank coefficient of $\sim-0.43$ seen in the data is very different to the rank coefficient of the simulation, with only $\sim$1\% of the realisations showing an anti-correlation stronger than that based on the original data (see Fig.~\ref{simulation_fcmdot}), so this is significant at close to $3\sigma$. 

Thus $f_c$ in our sample is not significantly correlated with $L_{\rm Bol}$, but it is with $\lambda_{\rm Edd}$ at $\sim$99\% significance. This indicates that changes in the covering factor are driven more by changes in the Eddington ratio, rather than by changes in the bolometric luminosity. This adds to a growing body of evidence that there are large-scale changes in the SED with $\lambda_{\rm Edd}$ \citep[][J12a, b]{2007MNRAS.381.1235V,2009MNRAS.392.1124V,2013ApJ...777...86L}. 
Therefore we find that the most basic of the unification models in which it is proposed that the observed AGN properties only depend on inclination are too simple, and there are changes in the shape of the SED which depend on $\lambda_{\rm Edd}$, as well as M$_{\rm BH}$ which sets the overall luminosity scale. However, the anti-correlation of the dust covering fraction with $\lambda_{\rm Edd}$ rather than $L_{\rm Bol}$ indicates a change in the larger scale geometry of the AGN rather than just the expected response of the dust to increasing illumination. Such a large-scale change may also be required to produce the observed anti-correlation of the forbidden series of the narrow emission lines with $\lambda_{\rm Edd}$, as Narrow Line Seyfert~1s and other high $\lambda_{\rm Edd}$ AGN are known to have weak [OIII] \citep[e.g. ][]{1992ApJ...397..442B,2016MNRAS.460.1716D}. Furthermore, \cite{2004ApJ...611..125L} speculate that this is due to the very inner regions of the accretion flow being progressively shielded by a wind, with increasing $\lambda_{\rm Edd}$. Thereby even if there is copious dust present, the irradiated fraction decreases as the ionising radiation becomes more collimated, and hence the reprocessed fraction drops. \cite{2008MNRAS.385L..43F} have discussed the fact that efficient coupling of dust to gas boosts the effect of radiation pressure feedback. The result is that absorbed AGN are mostly found at low Eddington ratios. Here, we are seeing a decrease of the (illuminated) dust fraction in type~1 AGN. The effect could be related to that noted by \cite{2008MNRAS.385L..43F} in absorbed AGN, with the feedback in our sample occurring out of our direct line of sight. 
Conversely, if the bulk of the MIR is emitted by dust located in the polar directions, then this result relates to the relative efficiency of illuminated dust emission in the line of sight.

\label{corr}
\begin{table}
\begin{center}
\caption{Spearman's correlation between different parameters}
\begin{tabular}{cccc}
\hline
\hline
Parameter 1 & Parameter 2 & $\rho_s$ & $d_s$ \\
\hline
$\lambda_{\rm Edd}$ & M$_{\rm BH}$ & -0.17 & 0.24 \\
$\lambda_{\rm Edd}$ & $L_{\rm Bol}$ & 0.61 & 1.64$\times10^{-06}$ \\
$\lambda_{\rm Edd}$ & $L_{\rm X-ray}$ & 0.28 & 0.04 \\
$\lambda_{\rm Edd}$ & $L_{\rm torus}$ & 0.439 & 0.001 \\
$\lambda_{\rm Edd}$ & $\kappa_{2-10}$ & 0.54 & 4.49$\times10^{-05}$ \\
$\lambda_{\rm Edd}$ & $R_{\rm cor}$ & -0.47 & 5.41$\times10^{-05}$ \\
$\lambda_{\rm Edd}$ & $f_{\rm pl}$ & -0.234 & 0.099 \\
$\lambda_{\rm Edd}$ & $\Gamma$ & 0.26 & 0.06 \\
$\lambda_{\rm Edd}$ & $f_c$ & -0.428 & 0.002 \\
$L_{\rm Bol}$ & $L_{\rm X-ray}$ & 0.76 & 7.8$\times10^{-11}$ \\
$L_{\rm Bol}$ & $L_{\rm torus}$ & 0.87 & 1.15$\times10^{-16}$ \\
$L_{\rm Bol}$ & $\kappa_{2-10}$ & 0.34 & 0.02 \\
$L_{\rm Bol}$ & $R_{\rm cor}$ & -0.368 & 0.008 \\
$L_{\rm Bol}$ & $f_{\rm pl}$ & -0.08 & 0.56 \\
$L_{\rm Bol}$ & $\Gamma$ & -0.08 & 0.57 \\
$L_{\rm Bol}$ & $f_c$ & -0.27 & 0.05 \\
M$_{\rm BH}$ & $L_{\rm Bol}$ & 0.61 & 2.16$\times10^{-06}$ \\
M$_{\rm BH}$ & $L_{\rm X-ray}$ & 0.69 & 1.71$\times10^{-08}$ \\
M$_{\rm BH}$ & $L_{\rm torus}$ & 0.65 & 2.66$\times10^{-07}$ \\
M$_{\rm BH}$ & $\kappa_{2-10}$ & -0.13 & 0.35 \\
M$_{\rm BH}$ & $R_{\rm cor}$ & 0.06 & 0.65 \\
M$_{\rm BH}$ & $f_{\rm pl}$ & 0.11 & 0.46 \\
M$_{\rm BH}$ & $\Gamma$ & -0.383 & 0.005 \\
M$_{\rm BH}$ & $f_c$ & 0.076 & 0.596 \\
$L_{\rm X-ray}$ & $L_{\rm torus}$ & 0.79 & 8.66$\times10^{-12}$ \\
$L_{\rm X-ray}$ & $\kappa_{2-10}$ & -0.29 & 0.04 \\
$L_{\rm X-ray}$ & $R_{\rm cor}$ & 0.05 & 0.71 \\
$L_{\rm X-ray}$ & $f_{\rm pl}$ & 0.27 & 0.06 \\
$L_{\rm X-ray}$ & $\Gamma$ & -0.34 & 0.01 \\
$L_{\rm X-ray}$ & $f_c$ & 0.007 & 0.96 \\
$L_{\rm torus}$ & $\kappa_{2-10}$ & 0.16 & 0.26 \\
$L_{\rm torus}$ & $R_{\rm cor}$ & -0.234 & 0.098 \\
$L_{\rm torus}$ & $f_{\rm pl}$ & 0.09 & 0.55 \\
$L_{\rm torus}$ & $\Gamma$ & -0.14 & 0.33 \\
$L_{\rm torus}$ & $f_c$ & 0.17 & 0.22 \\
$\kappa_{2-10}$ & $R_{\rm cor}$ & -0.65 & 2.31$\times10^{-07}$ \\
$\kappa_{2-10}$ & $f_{\rm pl}$ & -0.52 & 7.89$\times10^{-05}$ \\
$\kappa_{2-10}$ & $\Gamma$ & 0.434 & 0.001 \\
$\kappa_{2-10}$ & $f_c$ & -0.378 & 0.007 \\
$R_{\rm cor}$ & $f_{\rm pl}$ & -0.07 & 0.61 \\
$R_{\rm cor}$ & $\Gamma$ & -0.31 & 0.03 \\
$R_{\rm cor}$ & $f_c$ & 0.32 & 0.02 \\
$f_{\rm pl}$ & $\Gamma$ & -0.29 & 0.04 \\
$f_{\rm pl}$ & $f_c$ & 0.3 & 0.03 \\
$\Gamma$ & $f_c$ & -0.18 & 0.21 \\
\hline
\hline
\end{tabular}
\label{tab_corr}
\end{center}
\end{table}

\section{Conclusions}
\label{Sec6}

We present a detailed study of the dust covering factors for an X-ray/optically selected sample of unobscured type~1 AGN in the local Universe using the data available from \textit{XMM-Newton}, SDSS, \textit{WISE}, 2MASS, and UKIDSS. We used the method of SED modelling analysis to determine the covering factor of each source. Two important aspects of this work are that we have broadband spectra to determine $L_{\rm Bol}$, and a self-consistent model to estimate the contribution of the unobservable FUV region. We find a mean covering factor of $f_c$ = 0.30, with 0.02 $<f_c<$ 0.88 and a dispersion of the individual values of $\sigma_f$ = 0.17. The distribution shows a trend of anti-correlation with $\lambda_{\rm Edd}$ and $L_{\rm Bol}$, but further analysis based on simulations shows that only the trend with $\lambda_{\rm Edd}$ is significant at $\sim$99\%.  This implies a large-scale change in the geometry of the illuminated dust, rather than simply to a response from increasing $L_{\rm Bol}$. Division into sub-samples of radio-loud AGN and NLS1, do not reveal any significant differences in the distribution of covering factors from the whole sample. 
This argues against the presence of a strong additional driving parameter for $f_c$ in these sub-samples. However, the number of objects is small, and further studies are needed.

Our study is based on 51 sources, for which we have comprehensive multi-wavelength coverage. It would be valuable to extend this type of study to a larger sample of AGN, with a wider range of redshifts to test the correlations we find and the conclusions that we have drawn. It is relevant to select sources with different combinations of black hole mass and accretion rate since the behaviour of the accretion disc spectrum depends on these parameters. Further extension of this study can be expected from \textit{AstroSat} \citep{2014SPIE.9144E..1SS} which can obtain simultaneous observations in X-ray and UV bands.

\section*{Acknowledgements}

We thank the referee Konrad Tristram for his thorough review and useful comments.
We acknowledge the UGC-UKIERI Thematic Partnership 2015 (UGC 2014-15/02) for the support of the grant for this work. The first author is grateful to the Department of Science and Technology (No.SR/S2/HEP-07/2012) for the financial support. P.G. acknowledges the support of STFC (No. ST/J003697/2). CD acknowledges support under STFC grant ST/L00075X/1.

This work is based on observations obtained with \textit{XMM-Newton}, an ESA science mission with instruments and contributions directly funded by ESA Member States and NASA. This research has made use of the NASA/ IPAC Infrared Science Archive, which is operated by the Jet Propulsion Laboratory, California Institute of Technology, under contract with the National Aeronautics and Space Administration. This publication makes use of data products from the \textit{Wide-field Infrared Survey Explorer}, which is a joint project of the University of California, Los Angeles, and the Jet Propulsion Laboratory/California Institute of Technology, funded by the National Aeronautics and Space Administration. We have used the UKIDSS data from Data Release 10. The UKIDSS project is defined in \citep{2007MNRAS.379.1599L}. UKIDSS uses the UKIRT Wide Field Camera (WFCAM; \cite{2007A&A...467..777C}). The photometric system is described in \cite{2006MNRAS.367..454H}, and the calibration is described in \cite{2009MNRAS.394..675H}. The pipeline processing and science archive are described in Irwin et al (2009, in prep) and \cite{2008MNRAS.384..637H}. This publication makes use of data products from the Two Micron All Sky Survey, which is a joint project of the University of Massachusetts and the Infrared Processing and Analysis Center/California Institute of Technology, funded by the National Aeronautics and Space Administration and the National Science Foundation.

Funding for the SDSS and SDSS-II has been provided by the Alfred P. Sloan Foundation, the Participating Institutions, the National Science Foundation, the U.S. Department of Energy, the National Aeronautics and Space Administration, the Japanese Monbukagakusho, the Max Planck Society, and the Higher Education Funding Council for England. The SDSS Web Site is \url{http://www.sdss.org/}.

The SDSS is managed by the Astrophysical Research Consortium for the Participating Institutions. The Participating Institutions are the American Museum of Natural History, Astrophysical Institute Potsdam, University of Basel, University of Cambridge, Case Western Reserve University, University of Chicago, Drexel University, Fermilab, the Institute for Advanced Study, the Japan Participation Group, Johns Hopkins University, the Joint Institute for Nuclear Astrophysics, the Kavli Institute for Particle Astrophysics and Cosmology, the Korean Scientist Group, the Chinese Academy of Sciences (LAMOST), Los Alamos National Laboratory, the Max-Planck-Institute for Astronomy (MPIA), the Max-Planck-Institute for Astrophysics (MPA), New Mexico State University, Ohio State University, University of Pittsburgh, University of Portsmouth, Princeton University, the United States Naval Observatory, and the University of Washington.

This research has made use of the NASA/IPAC Extragalactic Database (NED) which is operated by the Jet Propulsion Laboratory, California Institute of Technology, under contract with the National Aeronautics and Space Administration.

\def\aj{AJ}%
\def\actaa{Acta Astron.}%
\def\araa{ARA\&A}%
\def\apj{ApJ}%
\def\apjl{ApJ}%
\def\apjs{ApJS}%
\def\ao{Appl.~Opt.}%
\def\apss{Ap\&SS}%
\def\aap{A\&A}%
\def\aapr{A\&A~Rev.}%
\def\aaps{A\&AS}%
\def\azh{AZh}%
\def\baas{BAAS}%
\def\bac{Bull. astr. Inst. Czechosl.}%
\def\caa{Chinese Astron. Astrophys.}%
\def\cjaa{Chinese J. Astron. Astrophys.}%
\def\icarus{Icarus}%
\def\jcap{J. Cosmology Astropart. Phys.}%
\def\jrasc{JRASC}%
\def\mnras{MNRAS}%
\def\memras{MmRAS}%
\def\na{New A}%
\def\nar{New A Rev.}%
\def\pasa{PASA}%
\def\pra{Phys.~Rev.~A}%
\def\prb{Phys.~Rev.~B}%
\def\prc{Phys.~Rev.~C}%
\def\prd{Phys.~Rev.~D}%
\def\pre{Phys.~Rev.~E}%
\def\prl{Phys.~Rev.~Lett.}%
\def\pasp{PASP}%
\def\pasj{PASJ}%
\def\qjras{QJRAS}
\def\rmxaa{Rev. Mexicana Astron. Astrofis.}%
\def\skytel{S\&T}%
\def\solphys{Sol.~Phys.}%
\def\sovast{Soviet~Ast.}%
\def\ssr{Space~Sci.~Rev.}%
\def\zap{ZAp}%
\def\nat{Nature}%
\def\iaucirc{IAU~Circ.}%
\def\aplett{Astrophys.~Lett.}%
\def\apspr{Astrophys.~Space~Phys.~Res.}%
\def\bain{Bull.~Astron.~Inst.~Netherlands}%
\def\fcp{Fund.~Cosmic~Phys.}%
\def\gca{Geochim.~Cosmochim.~Acta}%
\def\grl{Geophys.~Res.~Lett.}%
\def\jcp{J.~Chem.~Phys.}%
\def\jgr{J.~Geophys.~Res.}%
\def\jqsrt{J.~Quant.~Spec.~Radiat.~Transf.}%
\def\memsai{Mem.~Soc.~Astron.~Italiana}%
\def\nphysa{Nucl.~Phys.~A}%
\def\physrep{Phys.~Rep.}%
\def\physscr{Phys.~Scr}%
\def\planss{Planet.~Space~Sci.}%
\def\procspie{Proc.~SPIED}%
\let\astap=\aap
\let\apjlett=\apjl
\let\apjsupp=\apjs
\let\applopt=\ao
\bibliographystyle{mn2e} 

\appendix
\label{Appendix}

\section{IR SED templates}
\label{A}

\subsection{\textit{agndust} templates}

\begin{figure}
\begin{center}
\includegraphics[trim=0cm 0cm 0cm 0.0cm, clip=true, width =8.5cm, angle=0]
{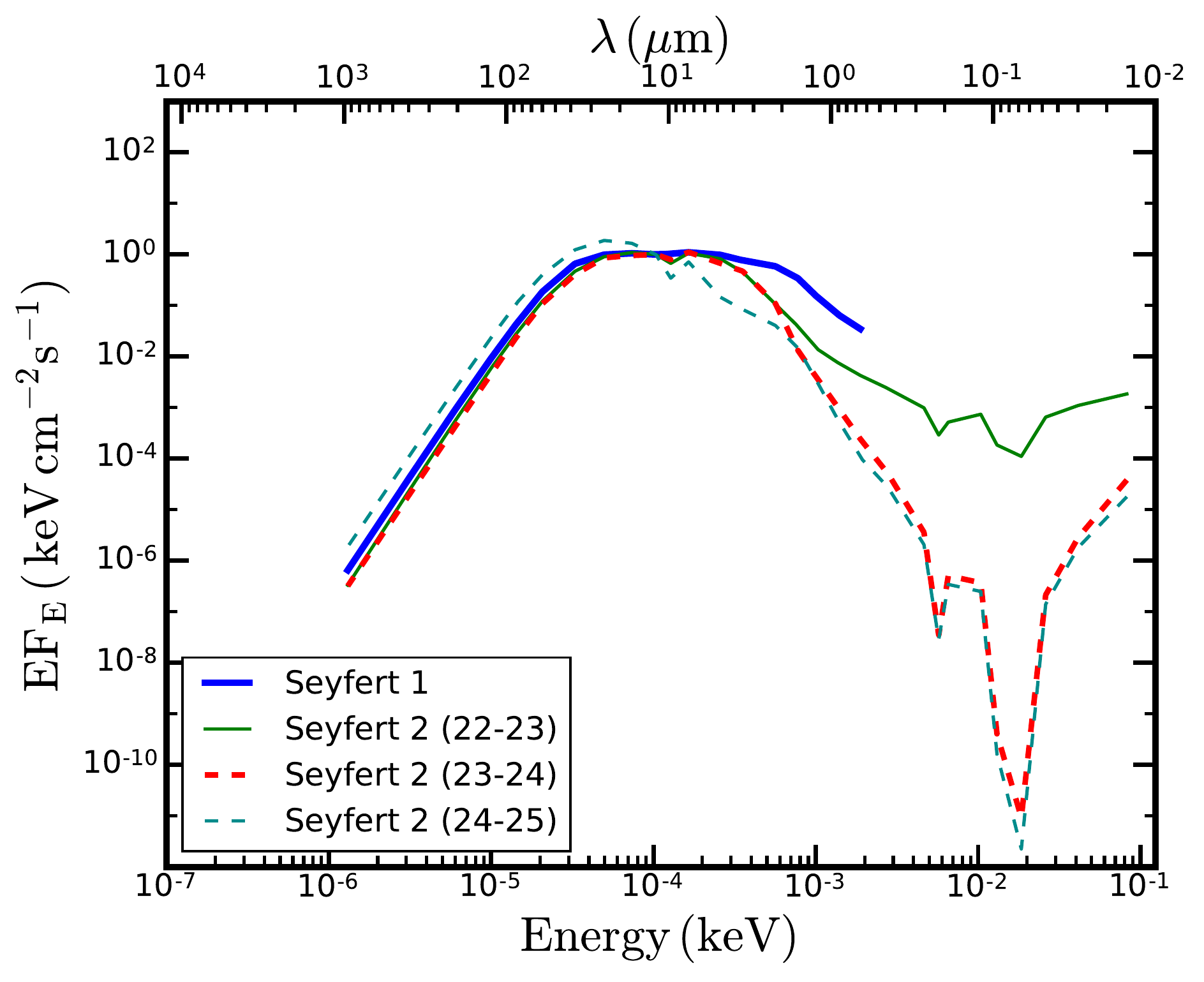}
\caption{\small The rest-frame Silva SED templates \citep{2004MNRAS.355..973S} for Seyfert~1 and Seyfert~2 galaxies, normalised at 12~$\mu$m. Seyfert~2 (22-23), Seyfert~2 (23-24) and Seyfert~2 (24-25) denote SEDs with 10$^{22}<N_H<$10$^{23}$~cm$^{-2}$, 10$^{23}<N_H<$10$^{24}$~cm$^{-2}$ and 10$^{24}<N_H<$10$^{25}$~ cm$^{-2}$, respectively. We isolate the infrared hump of the Seyfert~1 SED to concentrate on the dust reprocessing component of the torus.}
\label{Fig_silva}
\end{center}
\end{figure}

The model \textit{agndust} makes use of the templates from \cite{2004MNRAS.355..973S} 
who derived the nuclear infrared spectral energy distributions for a sample of obscured and unobscured Seyfert galaxies. They divided the observed IR SEDs into intervals of intrinsic absorbing column density $N_{\rm H}$. In order to exclude the objects with $N_{\rm H} >$~10$^{25}$~cm$^{-2}$ these SEDs were already normalised by the unabsorbed hard X$-$ray (2$-$10~keV band) flux. They obtained four different SEDs averaged within bins of absorbing $N_{\rm H}$. One among these nuclear IR SEDs corresponds to Seyfert~1 objects with $N_{\rm H}~<~10^{22}$~cm$^{-2}$ and the other three SEDs are for Seyfert~2 galaxies with $10^{22}<N_{\rm H}<10^{23}$~cm$^{-2}$, $10^{23}<N_{\rm H}<10^{24}$~cm$^{-2}$ and $10^{24}<N_{\rm H}<10^{25}$~ cm$^{-2}$. The \textit{agndust} model can make use of the four SEDs for modelling the IR data. In view of our sample selection, we make use of only the Seyfert~1 template ​by excluding the part of the SED at shorter wavelengths, ​plotted​​ in​ blue in Fig.~\ref{Fig_silva}.

\subsection{\textit{hostpol} templates}

\begin{figure}
\begin{center}
\includegraphics[trim=0cm 0cm 0cm 0.0cm, clip=true, width =8.5cm, angle=0]{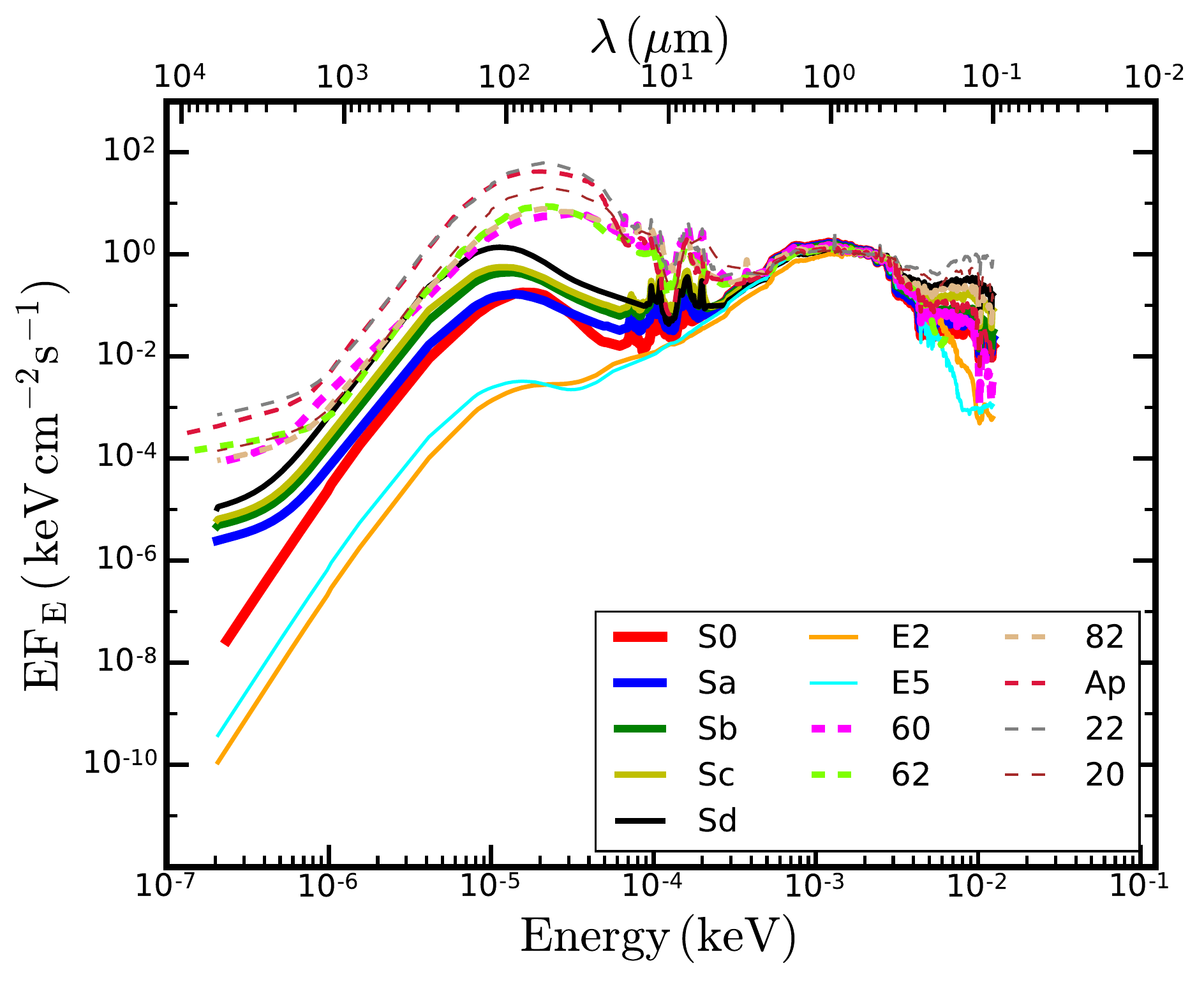}
\caption{\small The host galaxy SEDs from \citealt{2007ApJ...663...81P} (flux density normalised at 5500~\AA) for spirals (S0$-$Sd), ellipticals (E2 \& E5) and starburst galaxies. The starburst templates correspond to the SEDs of NGC~6090 (60), NGC~6240 (62), M~82 (82), Arp~220 (Ap), IRAS~22491-1808 (22), and IRAS~20551-4250~(20).}
\label{Fig_polletta}
\end{center}
\end{figure}

\begin{table}
\begin{center}
\caption{\textit{hostpol} model components and corresponding SED templates of host galaxies used in our analysis ($^*$starburst galaxies).}
\begin{tabular}{cc}
\hline
\textit{hostpol} component & SED template \\ 
\hline
\textit{host}01 & S0 \\ 
\textit{host}02 & Sa \\ 
\textit{host}03 & Sb \\ 
\textit{host}04 & Sc \\ 
\textit{host}05 & Sd \\ 
\textit{host}06 & E2 \\ 
\textit{host}07 & E5 \\ 
\textit{host}08 & NGC~6090$^*$ \\ 
\textit{host}09 & NGC~6240$^*$ \\ 
\textit{host}10 & M~82$^*$ \\ 
\textit{host}11 & Arp~220$^*$ \\ 
\textit{host}12 & IRAS~22491-1808$^*$ \\ 
\textit{host}13 & IRAS~20551-4250$^*$ \\ 
\hline 
\end{tabular}
\label{host}
\end{center}
\end{table} 

The \textit{hostpol} component uses the IR SED templates from SWIRE template library \citep{2007ApJ...663...81P}. The SWIRE template library has 25 IR SED templates which cover the wavelength range between 1000\AA~ and 1000$\mu$m. The library consists of 3 ellipticals, 7 spirals, 6 starbursts, 7 AGN and 2 composite (starburst+AGN) templates. The AGN templates comprise 3 type~1 and 4 type~2 AGN SEDs. The 13 SWIRE templates we used for modelling the host galaxy IR emission using \textit{hostpol} model (See Table~\ref{host}) are plotted in Fig.~\ref{Fig_polletta}.

\section{Broadband SED fits}
\label{B}
\begin{figure*}
\begin{center}
\caption{The broadband SED fitting plots for the 51 sources. We fit the absorbed SEDs to the observed data and the resultant best-fit models are shown in dashed grey line. However, we are mainly concerned with the measurements of $L_{\rm Bol}$, for which we illustrate the intrinsic model (solid red) together with the deabsorption corrections applied to the data. The individual model components {\sc optxagnf}, \textit{agndust} and \textit{hostpol} are plotted in dashed blue, solid green and solid yellow, respectively. Data from \textit{XMM-Newton} EPIC, Optical Monitor, SDSS, UKIDSS/2MASS, and \textit{WISE} are respectively represented by black dots, circles, diamonds, triangles, and squares. X-ray data have been rebinned for plotting purpose.}
\includegraphics[trim=0cm 0cm 0cm 0.0cm, clip=true,width=17cm, angle=0]
{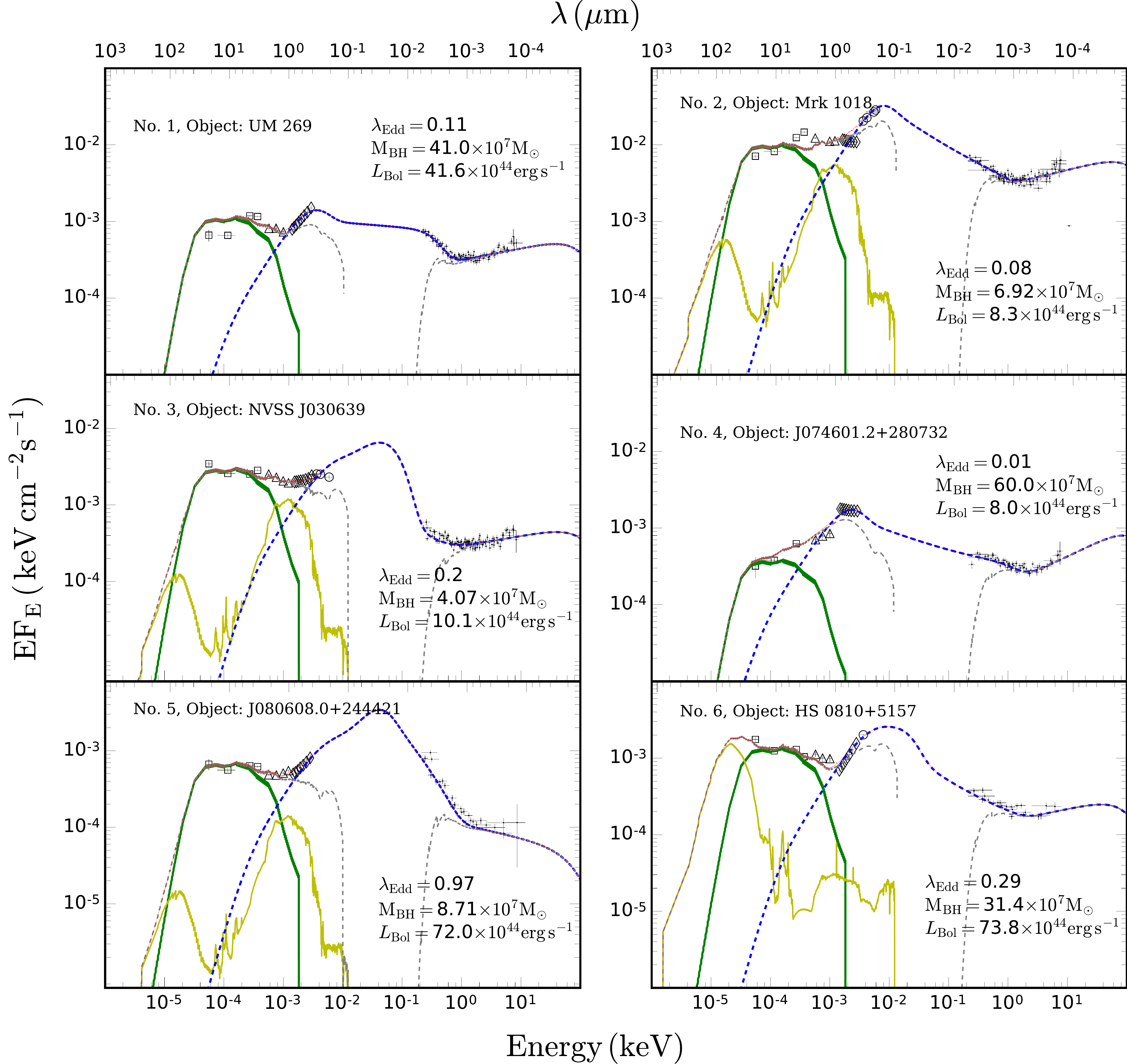}
\label{Fig_sed}
\end{center}
\end{figure*}

The broadband SED fits for the complete sample are shown in Fig.~\ref{Fig_sed}.

\onecolumn

\begin{figure*}
\begin{center}
\includegraphics[trim=0cm 0cm 0cm 0.0cm, clip=true,width=17.0cm, angle=0]
{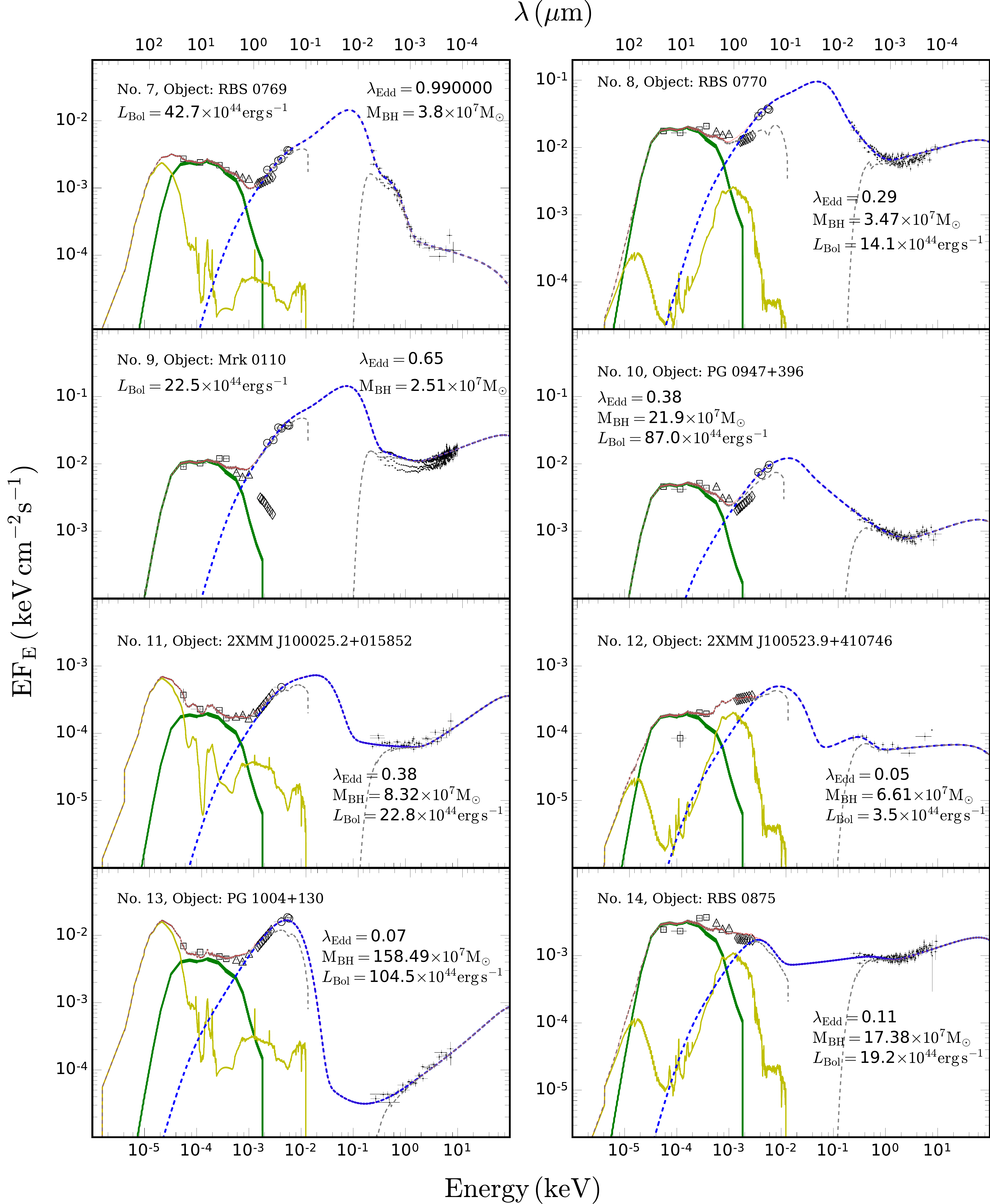}
\captionsetup{labelformat=empty}
\end{center}
\end{figure*}

\begin{figure*}
\begin{center}
\includegraphics[trim=0cm 0cm 0cm 0.0cm, clip=true,width=17.0cm, angle=0]
{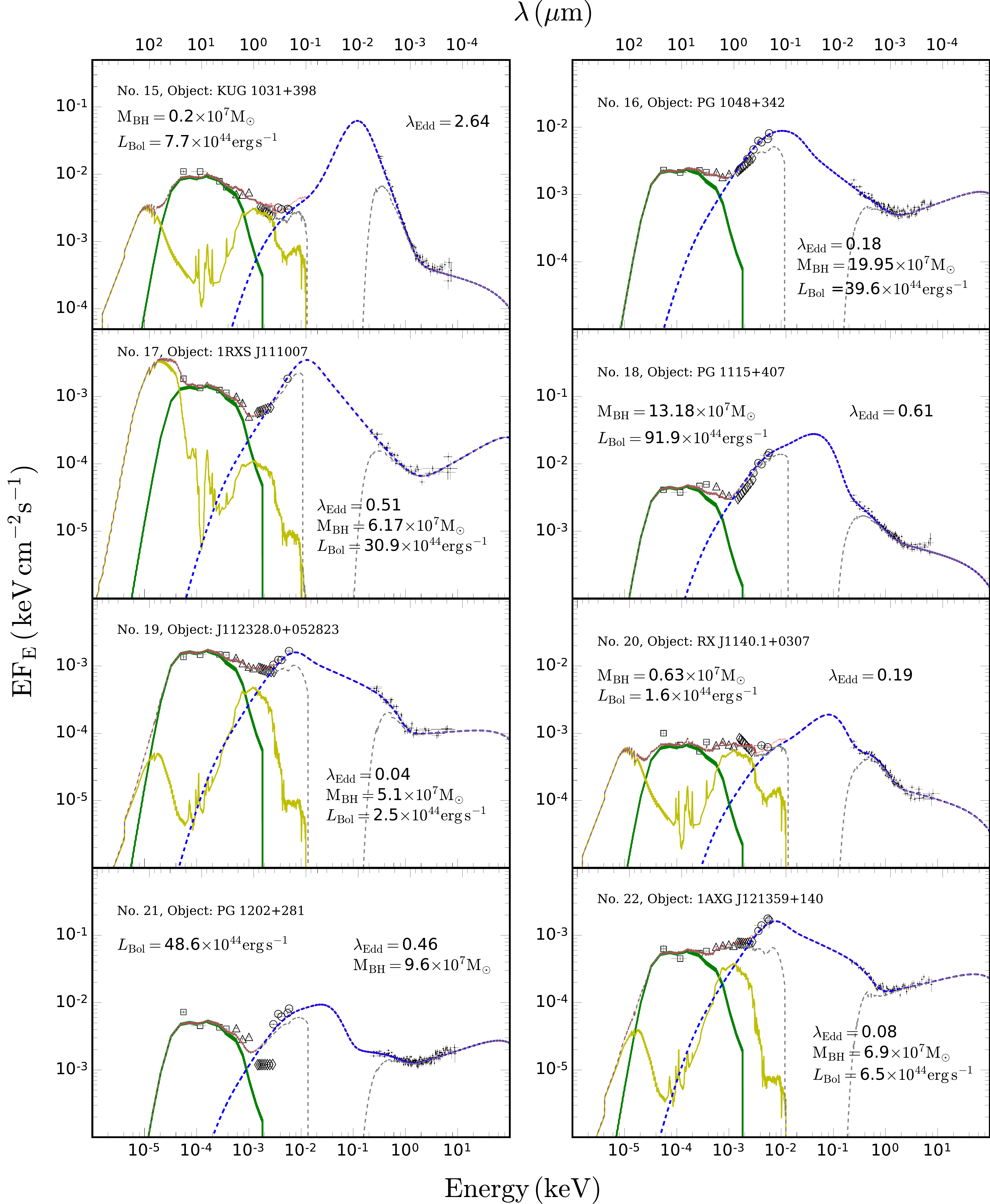}
\captionsetup{labelformat=empty}
\end{center}
\end{figure*}

\begin{figure*}
\begin{center}
\includegraphics[trim=0cm 0cm 0cm 0.0cm, clip=true,width=17.0cm, angle=0]
{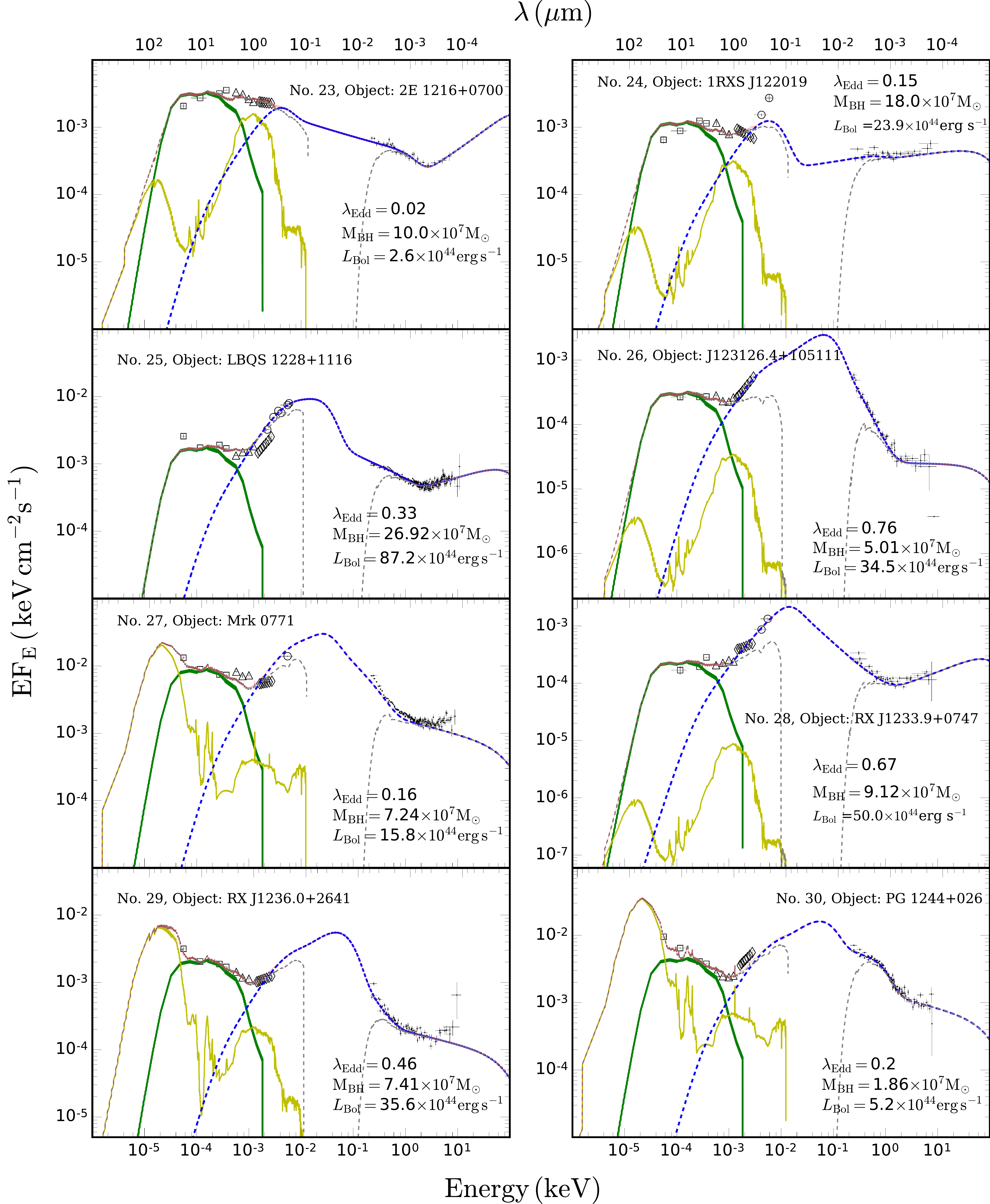}
\captionsetup{labelformat=empty}
\end{center}
\end{figure*}

\begin{figure*}
\begin{center}
\includegraphics[trim=0cm 0cm 0cm 0.0cm, clip=true,width=17.0cm, angle=0]
{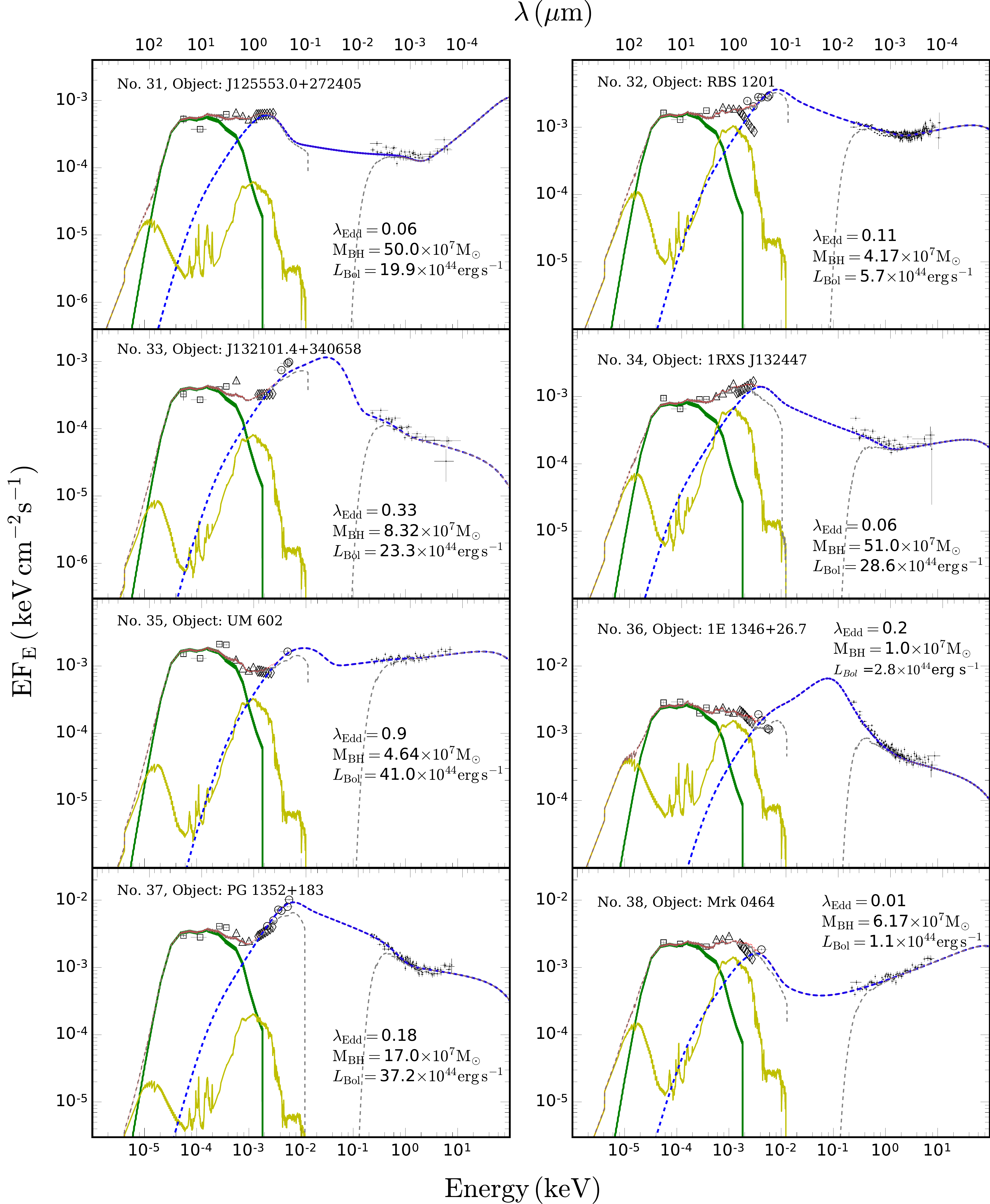}
\captionsetup{labelformat=empty}
\end{center}
\end{figure*}

\begin{figure*}
\begin{center}
\includegraphics[trim=0cm 0cm 0cm 0.0cm, clip=true,width=17.0cm, angle=0]
{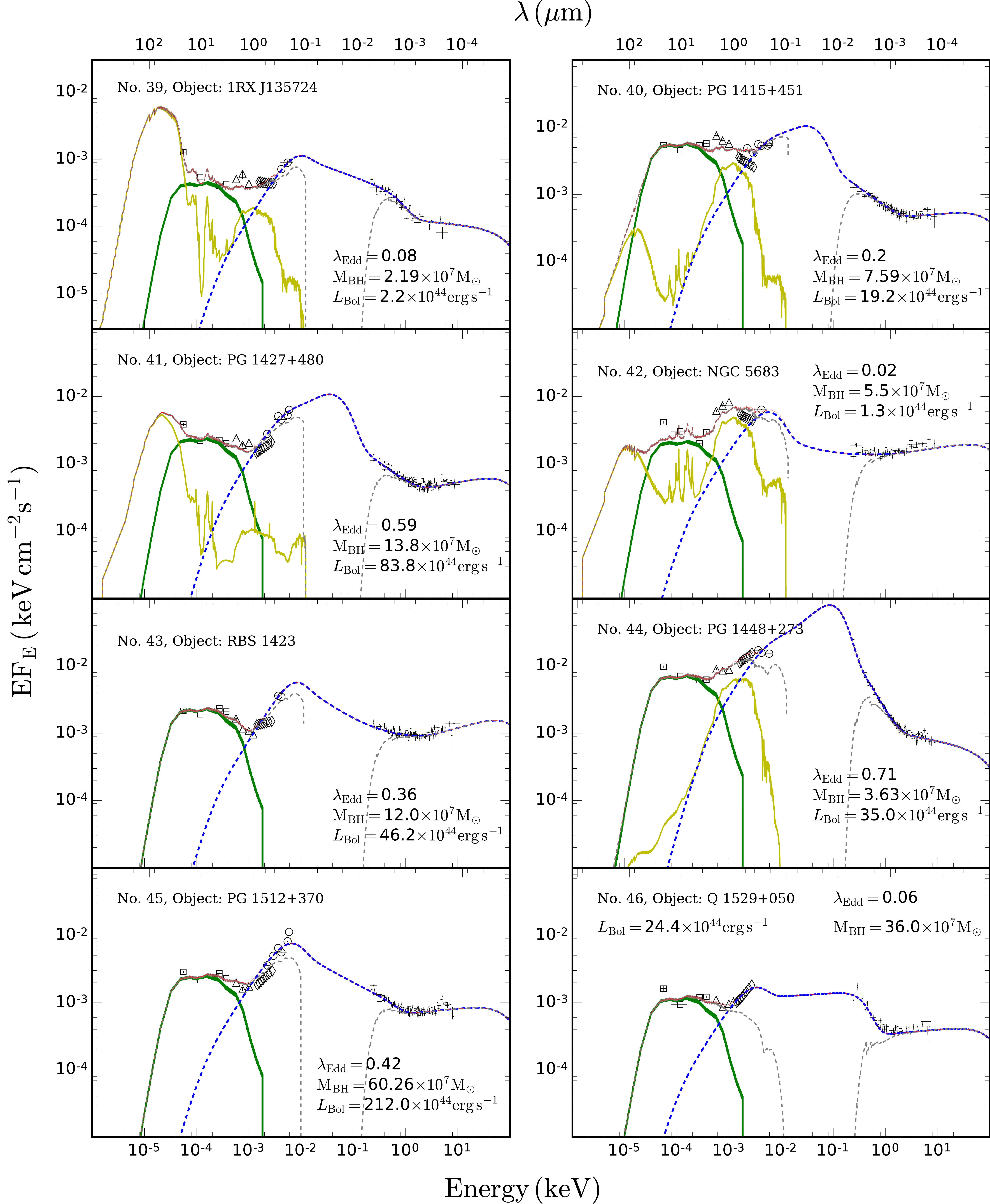}
\captionsetup{labelformat=empty}
\end{center}
\end{figure*}

\begin{figure*}
\begin{center}
\includegraphics[trim=0cm 0cm 0cm 0.0cm, clip=true,width=17.0cm, angle=0]
{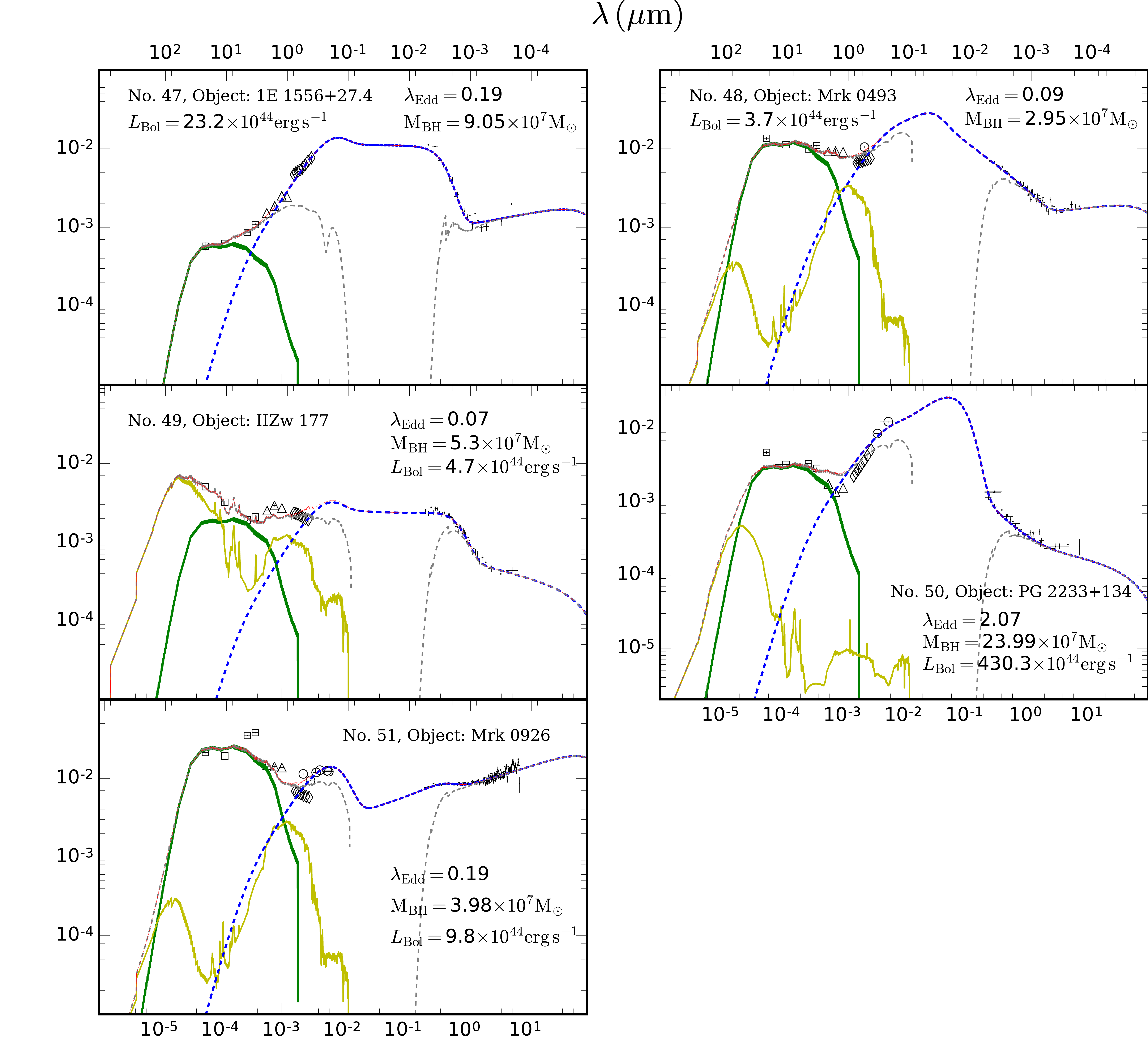}
\captionsetup{labelformat=empty}
\end{center}
\end{figure*}

\twocolumn

\section{Notes on individual sources}
\label{C}

\subsection{Discrepancies in SDSS data}

The SDSS data points in the SEDs of the objects Mrk~0110 (No.~9), PG~0947+396 (No.~10), PG~1202+281 (No.~21), LBQS~1228+1116 (No.~25), PG~1244+026 (No.~30), RBS~1201 (No.~32), PG~1415+451 (No.~40), NGC~5683 (No.~42), PG~1512+370 (No.~45) and PG~2233+134 (No.~50) show some clear deviations from the broadband continuum model. This may be attributed to the intrinsic variability of the sources. Since the observations of SDSS and OM are not simultaneous, there may be some discrepancy between the two data sets. The SED of Mrk~0110 shows large offset in the SDSS data points. J12a discussed this source and mentioned that it is an extreme example of this behaviour. In order to check the influence of the discrepancy, we fit these sources with the same model without using the SDSS data. We observed that $f_c$ remain unchanged in 8 sources, and in only two cases (PG~0947+396 (No.~10) \& PG~1202+281 (No.~21)), there is a drop by a factor of $\sim$2.

\subsection{Sources with multiple \textit{hostpol} templates}

In the case of Mrk 0110 (No.~9) and PG 1115+407 (No.~18), the normalisations of all 
\textit{hostpol} templates are nearly zero and are equally fitted by multiple host galaxy templates. For Mrk~0110, all the \textit{hostpol} templates provide the same fit-statistic and even the same spectral parameters. However, Mrk~0110 is identified as an Sa galaxy \citep{2012AstL...38..475K}. Hence, in the main text, we mentioned the SED of spiral galaxy type-a (\textit{host}02) as the best-fit \textit{hostpol} model for this object. In the case of PG 1115+407, more than one \textit{hostpol} template gave the same $\chi^2$ and slightly different parameter values. However, the morphological type of the source is unknown. In this case, we have adopted S0 template as the best-fit component.

\subsection{Super-Eddington sources}

There are two sources in our sample, KUG~1031+398 (No.~15) and PG~2233+134 (No.~50), which have super-Eddington accretion rates. Among these KUG~1031+398 has the highest value of $\lambda_{\rm Edd}$ ($\sim$2.64) and lowest black hole mass, whereas PG~2233+134 has the highest value of bolometric luminosity. In these sources, we attempt to fit the data by fixing $\lambda_{\rm Edd}$ to 1 and letting M$_{\rm BH}$ be a free parameter. For KUG~1031+398 this resulted in an improvement in the fit ($\Delta\chi^2$ = -58.5) for a change in M$_{\rm BH}$ from 1.7$\times$10$^6M_\odot$ to 3.8$\times$10$^6M_\odot$. Here, the covering factor increased from 0.23 to 0.34. As discussed by \cite{2016MNRAS.455..691J} this may be a super-Eddington source and the black hole mass obtained from the SED-fitting may not be correct. In such sources, the super-Eddington flow may not be well fit by a model that conserves energy. In the case of PG~2233+134, M$_{\rm BH}$ remained unchanged while $\chi^2$(/dof) increased from 893.6(/173) to 2641.2(/173). In this case also, $f_c$ shows an increase from 0.11 to 0.22. Although the inclination effects are not taken into account, it seems likely that this AGN is indeed a super-Eddington source. It is perhaps surprising that we are able to fit it as a super-Eddington source since such high Eddington sources probably power strong winds. So energy conservation is not appropriate due to loss of radiative power to the wind.

\section{Other Torus templates}
\label{D}

\cite{2011MNRAS.414.1082M} (hereafter M11) have constructed a range of intrinsic MIR to FIR (6-100~$\mu m$) SEDs of a sample of X-ray selected local AGN with moderate luminosities. We have fitted our sample with M11 SED by extending it down to about 0.6~$\mu m$ to match the range of our \textit{agndust} \citep{2004MNRAS.355..973S} template for Seyfert~1. The covering factors obtained with this template have a very similar range, mean and scatter as that for \textit{agndust}. Although the SED fits and spectral parameters are comparable to that of \textit{agndust}, for most of the sources the fit resulted in comparatively poor $\chi^2$. This is probably due to fact that the extended M11 SED is narrow and so fails to cover the peak in emission around the NIR region.

We have also attempted fitting the data with Clumpy SED for type~1 AGN \citep{2010A&A...523A..27H} with inclination 30~degrees, by modifying the template to match the wavelength range of \textit{agndust}. But again we note that the template is narrow compared to \textit{agndust} SED, and peaks around 10$\mu$m. Also, the torus luminosity for this component over 1$-$1000~$\mu$m is a factor of $\sim$1.3 lower than that for \textit{agndust} SED. The fits provide poor statistic for $\sim$70\% of the sources in the sample and a lower range of covering factors $\sim0.01-0.5$ with a mean around 0.18.

The data appear to require an extra hot dust component in the NIR, not covered by the above templates. We tested this for UM~269 (No.~1) by adding a black body component (XSPEC model \textit{bbody}) with temperature $\sim$1000$-$1500~K. For this source, a black body with $\sim$1000~K temperature in the NIR region, in addition to the Clumpy SED provided a better $\chi^2$ and resulted in a similar covering factor ($\sim$0.31) as we obtained when using \textit{agndust}. Hence the overall results are consistent with those obtained before.

\begin{figure*}
\begin{center}
\includegraphics[trim=0cm 0cm 0cm 0.0cm, clip=true,width=8.0cm, angle=0]
{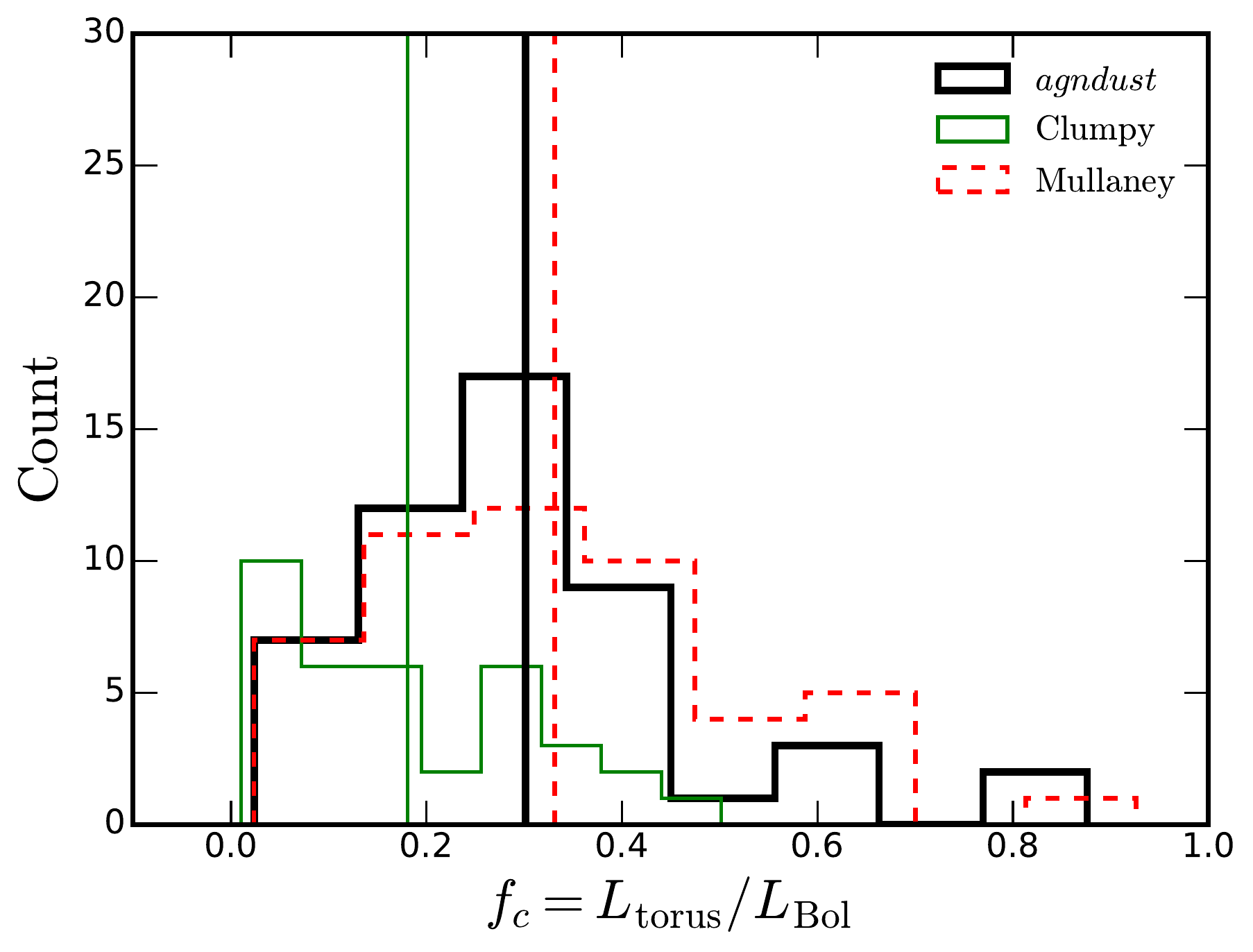}
\caption{\small Histograms of covering factors for \textit{agndust} SED (thick black line), extended M11 SED (dashed red line) and Clumpy SED for type~1 AGN (thin green line). The vertical lines represent the mean values of $f_c$ for different distributions.}
\label{otherSEDs}
\end{center}
\end{figure*}

The histograms of covering factors obtained for \textit{agndust}, extended M11 and Clumpy SEDs are shown in Fig.~\ref{otherSEDs}.

\section{Analysis with 2MASS data}
\label{E}

\begin{figure*}
\begin{center}
\includegraphics[trim=0cm 0cm 0cm 0.0cm, clip=true, width =8.0cm, angle=0]
{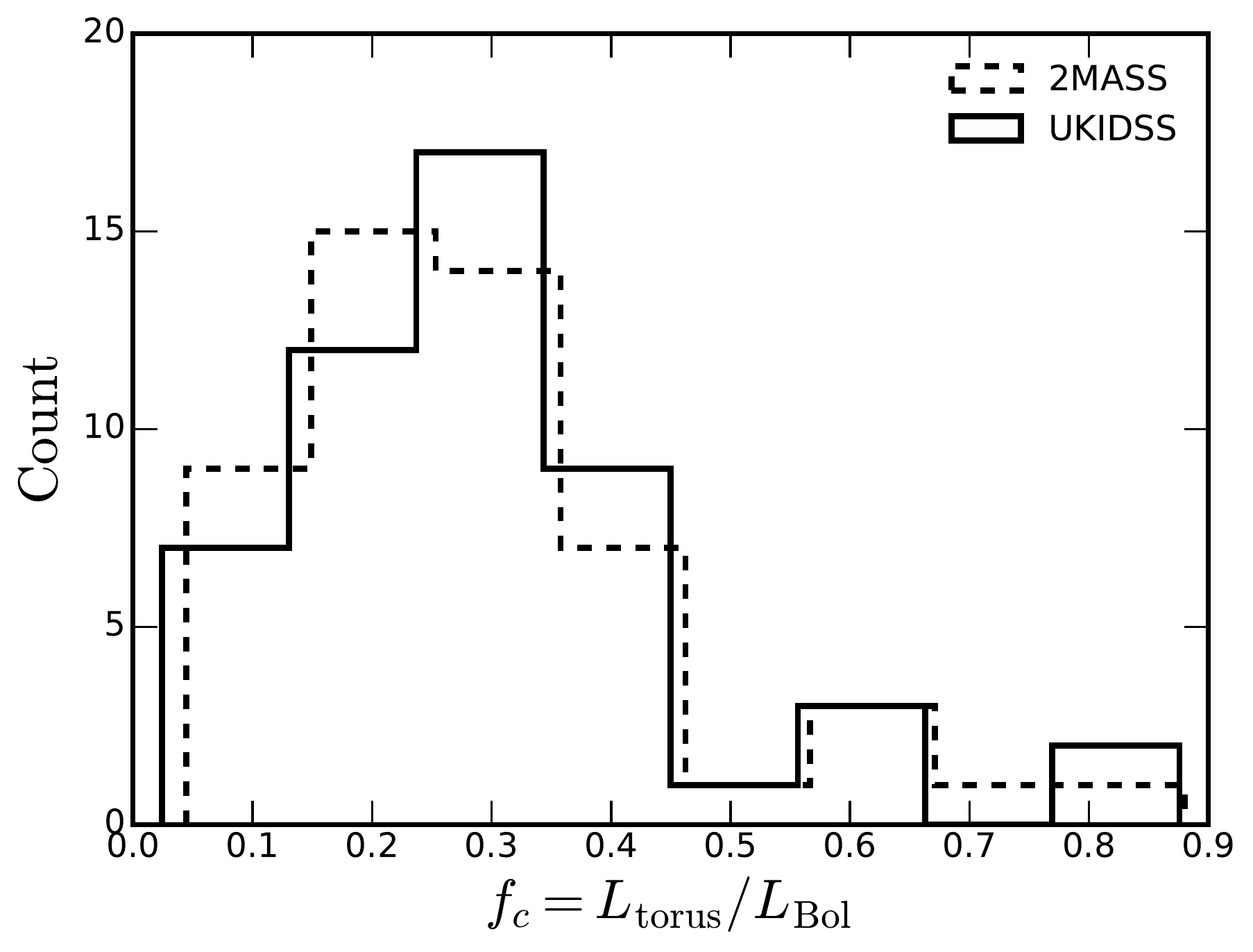}
\caption{\small Distributions of covering factors obtained using 2MASS data (dashed line) and UKIDSS data (solid line) for the NIR band.}
\label{fc_2m}
\end{center}
\end{figure*}

As mentioned in Section~\ref{IRdata} we opted to use NIR data from UKIDSS instead of 2MASS because of the smaller aperture size of the UKIDSS camera (2\arcsec diameter for UKIDSS \& 4\arcsec radius for 2MASS), which therefore reduces the contribution from the host galaxy. Also, three sources in our sample lack 2MASS data and in those cases we require UKIDSS data. However, for comparison, we have analysed the entire sample by fitting the broadband SED with 2MASS data (if available) for the NIR band. We find that the distribution of covering factors is similar to that obtained when using the UKIDSS data. Both distributions of $f_c$ have the same mean and standard deviation. The histograms of $f_c$ are shown in Figure~\ref{fc_2m}. We have also checked the correlation of $f_c$ with $L_{\rm Bol}$ and $\lambda_{\rm Edd}$ obtained for 2MASS data. The trend of $f_c$ between these parameters is essentially the same as we found when using the UKIDSS data.

\end{document}